\documentclass[prb, aps, twocolumn, footinbib, groupedaddress, showpacs,
citeautoscript]{revtex4-1}
\usepackage{amssymb}
\usepackage{amsmath}
\usepackage{amsfonts}
\DeclareMathOperator{\trace}{Tr}
\usepackage{color}
\usepackage{graphicx}
\usepackage{bm}
\newcommand* {\vek}[1]{{\ensuremath{\bm{\mathrm{#1}}}}}
\newcommand* {\vekc}[1]{{\ensuremath{\bm{\mathcal{#1}}}}}
\newcommand* {\kk}{\vek{k}}
\newcommand* {\rr}{\vek{r}}
\newcommand* {\frack}[2]{{\Ts\frac{#1}{#2}}}
\newcommand* {\Ds}{\displaystyle}
\newcommand* {\Ts}{\textstyle}

\newcommand* {\ket}[1]{\ensuremath{| {#1} \rangle}}
\newcommand* {\braket}[1]{\ensuremath{\langle {#1} \rangle}}
\usepackage{array}
\newcolumntype {s}[1]{@{\hspace{#1}}} 
\newcolumntype {R}{>{$}r<{$}}         
\newcolumntype {C}{>{$}c<{$}}         
\newcolumntype {L}{>{$}l<{$}}         

\newcommand* {\kcomp}{\kappa}
\newcommand* {\kvek}{\bm{\kcomp}}
\newcommand* {\tvek}[2][c]{\left( \begin{array}{s{0.15em}#1s{0.15em}}
     #2\end{array} \right)}
\newcommand* {\strain}{\epsilon}
\newcommand* {\Strain}{\vek{\epsilon}}
\newcommand* {\Ee}{\mathcal{E}}
\newcommand* {\koeff}[3]{#1^{#2}_{#3}}
\newcommand* {\tgamma}{\tilde{\gamma}}
\newcommand* {\etal}{\textit{et al.}}

\newcommand{\allowed}[1]{\mbox{$\bm{#1}$}}
\newcommand{\allowednew}[1]{\allowed{#1}\ensuremath{\ddagger}}
\newcommand{\forbidden}[1]{\mbox{$#1$}}

\newcommand{\num}[3][]{#2~#1\mathrm{#3}}
\newcommand{\nume}[4][]{#2 \times 10^{#3}~#1\mathrm{#4}}

\begin{document}

\title{Electromagnetic coupling of spins and pseudospins in bilayer graphene}

\author{R. Winkler}
\affiliation{Department of Physics, Northern Illinois University,
DeKalb, Illinois 60115, USA}
\affiliation{Materials Science Division, Argonne National
Laboratory, Argonne, Illinois 60439, USA}
\email{rwinkler@niu.edu}

\author{U. Z\"ulicke}
\affiliation{School of Chemical and Physical Sciences and
MacDiarmid Institute for Advanced Materials and Nanotechnology,
Victoria University of Wellington, PO Box 600, Wellington 6140,
New Zealand}
\email{uli.zuelicke@vuw.ac.nz}

\date{June 30, 2015}

\begin{abstract}
  We present a detailed theoretical study of bilayer-graphene's
  electronic properties in the presence of electric and magnetic
  fields.  Using group-theoretical methods, we derive an invariant
  expansion of the Hamiltonian for electron states near the
  $\vek{K}$ point of the Brillouin zone. In contrast to known
  materials, including single-layer graphene, any possible coupling
  of physical quantities to components of the external
  \emph{electric\/} (magnetic) field has a counterpart where the
  analogous component of the \emph{magnetic\/} (electric) field
  couples to exactly the same combination of quantities.  For
  example, a purely electric spin splitting appears as the
  magneto-electric analogue of the familiar magnetic Zeeman spin
  splitting.  The measurable thermodynamic response induced by
  magnetic and electric fields is thus completely symmetric.  The
  Pauli magnetization induced by a magnetic field takes exactly the
  same functional form as the polarization induced by an electric
  field.  Our findings thus reveal unconventional behavior of spin
  and pseudospin degrees of freedom in their coupling to external fields.
  We explain how these counterintuitive couplings are
  consistent with fundamental principles such as time reversal symmetry.  For
  example, only a magnetic field can give rise to a macroscopic spin
  polarization, whereas only a perpendicular electric field can
  induce a macroscopic polarization of the sublattice-related
  pseudospin degree of freedom characterizing the intravalley
  orbital motion in bilayer graphene.  These rules enforced by
  symmetry for the matter-field interactions clarify the nature of
  spins versus pseudospins.  We also provide numerical values of
  prefactors for relevant coupling terms.  While our theoretical
  arguments use bilayer graphene as an example, they are generally
  valid for any material with similar symmetries.  The unusual
  equivalence of magnetic and electric fields discussed here can
  provide the basis for designing more versatile device
  architectures for creating polarizations and manipulating the
  orientation of spins and pseudospins.
\end{abstract}


\maketitle

\section{Introduction}

It is normally the case that physical effects associated with
electric fields are qualitatively different from those associated
with magnetic fields. The distinct physics related with the two
types of fields is generally mandated by their opposite behavior
under symmetry transformations: an electric field $\vek{\Ee}$
(magnetic field $\vek{B}$) is odd (even) under spatial inversion and
even (odd) under time reversal.  However, in certain
materials~\cite{heh08, ess09, ess10}, the clear separation between
electric and magnetic effects turns out to be blurred because time
reversal and/or inversion symmetries are broken (e.g., in
multiferroics~\cite{spa05, fie05, ram07}), or because the material's
band structure exhibits a special topological
structure~\cite{qi08}. In these \emph{magneto-electric\/}
media~\cite{ode70}, \emph{orbital\/} magnetic polarizations can be
induced by electric fields (and \textit{vice versa\/}) in a way
which realizes a condensed-matter physics analog of axion
electrodynamics~\cite{wil87, fra08}. Here we show that electrons at
a Dirac point in bilayer graphene (BLG) experience a previously
unknown type of electromagnetism in which the equivalence between
electric and magnetic effects is virtually complete: every coupling
of an electron's degrees of freedom to a \emph{magnetic\/} field is
matched by an analogous coupling of the \emph{same} degrees of
freedom to an \emph{electric\/} field. This unusual duality of
matter-field interactions is the physical origin of a recently
predicted magneto-electric response in valley-isospin-polarized
BLG~\cite{zue14}, but its implications are much broader as we show
in this paper.

\begin{figure}[bp]
  \includegraphics[width=1.0\columnwidth]{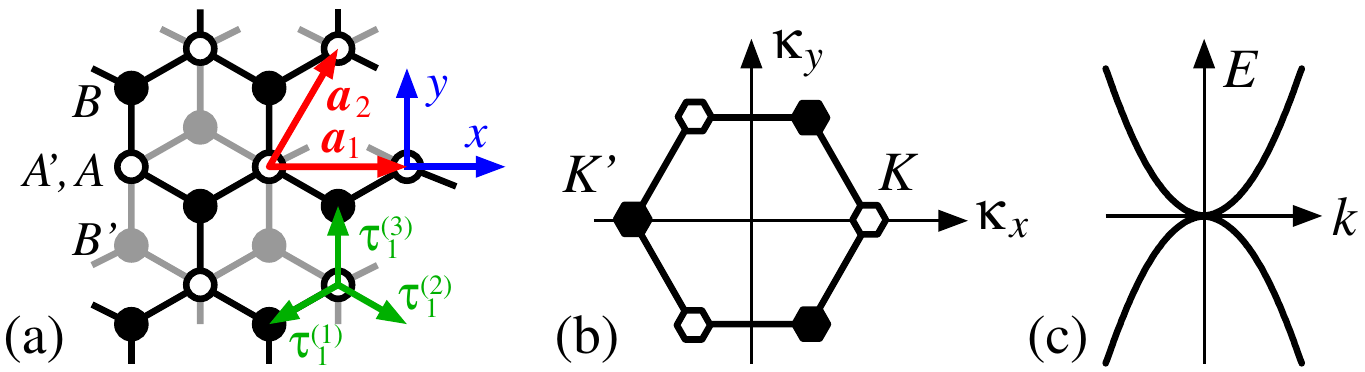}
  \caption{\label{fig:lattice} (a) Crystal structure of bilayer
  graphene. The honeycomb structure of the upper (lower) layer is
  marked in black (gray).  Atoms in sublattices $A$ and $A'$ ($B$
  and $B'$) are marked with open (closed) circles. (b) Brillouin
  zone and its two inequivalent corner points $\vek{K}$ and
  $\vek{K}'$. The remaining corners are related with $\vek{K}$ or
  $\vek{K}'$ by reciprocal lattice vectors. (c) Dispersion $E(k)$
  near the $\vek{K}$ point. We have $\kk \equiv \kvek - \vek{K}$.}
\end{figure}

The crystal structure and Brillouin zone of BLG are illustrated in
Fig.~\ref{fig:lattice}(a) and (b), respectively. The band structure
near the $\vek{K}$ point is described by the effective
Hamiltonian~\cite{mcc06, cas09, mcc13}
\begin{equation}
  \label{eq:eff_BLG}
  \begin{array}[b]{r>{\displaystyle}l}
(\mathcal{H}^\vek{K}_\kk)_{33} (\kk) = & 
\frac{\hbar^2}{2 m_0}\left[ -u
\left(k_+^2 \, \sigma_+ + k_-^2 \, \sigma_- \right) + w\, k^2
\sigma_0  \right] \\[1.5ex] & \hspace{6em}
 - \, \hbar v \left( k_- \, \sigma_+ + k_+ \, \sigma_- \right) \; ,
  \end{array}
\end{equation}
where $\hbar$ is Planck's constant, $m_0$ is the free-electron mass,
$\vek{k}\equiv(k_x, k_y)$ is the electrons' wave vector measured
from $\vek{K}$, and the Pauli matrices $\sigma_{x,y,z}$ are
associated with the sublattice (or, equivalently, the layer-index)
pseudospin degree of freedom \cite{cas09}. In our notation,
$\sigma_0$ is the $2\times 2$ unit matrix,
$\sigma_\pm = (\sigma_x \pm i\sigma_y)/2$, and $k_\pm=k_x\pm i k_y$.
Numerical values for the (positive and dimensionless) prefactors
$u$, $w$ and the speed $v$ are well known \cite{mcc06, cas09}, see
below. Very close to the $\vek{K}$ point, the energy dispersion
resulting from Eq.\ (\ref{eq:eff_BLG}) mimics that of massless Dirac
electrons, as is the case in single-layer graphene (SLG). However,
as $u \gg w$, the dominant behavior of electrons in BLG is captured
by the quadratic dispersion shown in Fig.~\ref{fig:lattice}(c).

External fields turn out to have a great influence on the electronic
properties of charge carriers in BLG.  Previously, only the effects
of electric and magnetic fields directed perpendicular to the BLG
sheet have been considered, which can be described by the extended
effective Hamiltonian
\begin{equation}\label{eq:fieldHam}
  \begin{array}[b]{l} \mathcal{H}^\vek{K}_{33} (\kk, \Ee_z, B_z) =
(\mathcal{H}^\vek{K}_\kk)_{33} (\kk) \\[0.5ex] \hspace{2em} \Ds {} +
\frac{g_\mathrm{e}}{2}\, \frac{\Ee_z}{c}\, \mu_\mathrm{B}\, \sigma_z -
\frac{g_\mathrm{m}}{2}\, B_z\, \mu_\mathrm{B}\, \sigma_z +
e \, \xi_z\, \Ee_z B_z\, \sigma_0 \, . \hspace{1em}
  \end{array}
\end{equation} 
Here we have followed the common practice \cite{kit63} that
$\hbar\kk$ denotes both crystal momentum and, for a magnetic field
$B >0$ the operator of kinetic momentum,
$\hbar\kk = -i\hbar\nabla + e\vek{A}$, the components of which obey
the commutator relation $[k_x, k_y] = (e/i\hbar)B_z$.  According to
Eq.\ (\ref{eq:fieldHam}), (i)~a potential difference between the two
layers (equivalent to a finite electric field $\Ee_z$) opens up a
pseudospin gap~\cite{mcc06, mcc06b, oht06, zha09}
$g_\mathrm{e} \,\mu_\mathrm{B}\, \Ee_z/c$, (ii)~a magnetic field
$B_z$ induces a pseudospin Zeeman splitting~\cite{zha11}
$g_\mathrm{m} \,\mu_\mathrm{B}\,B_z$, and (iii)~the simultaneous
presence of fields $\Ee_z$ and $B_z$ leads to a (valley-contrasting,
see below) overall energy shift \cite{xia07, nak09, kos10, zha11,
zue14}, $e\,\xi_z\, \Ee_z B_z$.  In Eq.~(\ref{eq:fieldHam}), the
effective \textit{g} factors $g_\mathrm{e}$ and $g_\mathrm{m}$ as
well as the prefactor $\xi_z$ are material parameters,
$\mu_\mathrm{B}$ is the Bohr magneton, and $\vek{A}$ is the
electromagnetic vector potential satisfying
$(\vek{\nabla}\times\vek{A})_z = B_z$. The matter-field interactions
(i)--(iii) generate sizable effects for typical values of $\Ee_z$
and $B_z$.  As discussed in more detail below, we have
$g_\mathrm{e} \simeq 500$, $g_\mathrm{m} \simeq 6.2$, and
$\xi_z = 3 \times 10^{-4}~\mathrm{nm/T}$.

Inspection of Eq.~(\ref{eq:fieldHam}) reveals a surprising feature:
disregarding constant prefactors, the electron's interaction with
fields $\Ee_z$ and $B_z$ is symmetric with respect to the
interchange of $\Ee_z$ and $B_z$. Indeed, this observation is not a
coincidence. It reflects the unusual property of electron states
near the $\vek{K}$ point of the BLG Brillouin zone that crystal
symmetry does not distinguish between polar vectors such as the
electric field $\vekc{E}$ and axial vectors such as the magnetic
field $\vek{B}$. Moreover, the familiar constraints due to
time-reversal invariance are modified at the BLG $\vek{K}$ point
such that symmetry under time reversal likewise permits that
$\vekc{E}$ and $\vek{B}$ become interchangeable.

In the following, we provide a rigorous derivation of the
magneto-electric equivalence exhibited in BLG and discuss physical
ramifications, with most of our major findings given in
Secs.~\ref{sec:MEanal} and \ref{sec:loewdin}.  Noteworthy results
include Table~\ref{tab:invariants}, which juxtaposes lowest-order
$\vek{B}$- and $\vekc{E}$-dependent terms in the effective
Hamiltonian for electrons in BLG related by magneto-electric
equivalence and elucidates their opposite symmetry with respect to
the valley degree of freedom; Fig.~\ref{fig:visual}, which
illustrates and compares the spin and pseudospin textures induced by
each of these terms in the two valleys; and Table~\ref{tab:prefact},
which provides parametric expressions and numerical values for
prefactors.  In addition, we discuss the measurable thermodynamic
response induced by magnetic and electric fields in BLG, which turns
out to be completely symmetric under exchange of $\vek{B}$ and
$\vekc{E}$, see Eqs.~(\ref{eq:response}) and (\ref{eq:mag-pol}).
These features reveal unconventional behavior of spin and pseudospin
degrees of freedom induced by external fields.  Although seemingly
counterintuitive, our findings are consistent with fundamental
principles such as time reversal symmetry.  This is indicated, e.g.,
by the fact that only a magnetic field $B_z$ can give rise to a
macroscopic polarization of the real spin $s_z$, whereas only an
electric field $\Ee_z$ can induce a macroscopic polarization of the
pseudospin $\sigma_z$.  These symmetry-enforced rules for the
matter-field interactions can serve to distinguish the physical
nature of spins versus pseudospins.

Our paper is organized as follows.  To establish some important
notations and conventions, we start in Sec.~\ref{sec:tb} with a
brief review of results previously obtained for BLG within a
tight-binding analysis. The basic formalism and results from the
invariant expansion for the BLG band structure near the $\vek{K}$
point are given in Sec.~\ref{sec:invExp}. It reveals the
magneto-electric equivalence, which is analyzed in greater detail in
Sec.~\ref{sec:MEanal}.  To develop a more quantitative description
of the predicted effects, we apply the theory of invariants in
Sec.~\ref{sec:invar-swm} to extend the tight-binding model from
Sec.~\ref{sec:tb} to include spin-orbit coupling and the effect of
external magnetic and electric fields.  This extended model is then
analyzed in Sec.~\ref{sec:loewdin} by means of L\"owdin partitioning
in order to derive explicit expressions for the prefactors of the
terms describing the magneto-electric equivalence in BLG.  We finish
by drawing conclusions in Sec.~\ref{sec:outlook}.

\section{Review of Tight-Binding Analysis}
\label{sec:tb}

In this section we present a brief review of results previously
obtained for BLG within a tight-binding analysis.  This enables us
to establish some important notations and conventions.  Also, it
allows us to properly relate our findings to previous work.

The crystal structure of BLG is sketched in
Fig.~\ref{fig:lattice}(a). For definiteness, we use the basis
vectors in real space
\begin{equation}
  \label{eq:basis-vec}
  \vek{a}_1 = a \tvek{1 \\ 0 \\ 0} ,
  \qquad
  \vek{a}_2 = a \tvek{1/2 \\ \sqrt{3}/2 \\ 0} ,
  \qquad
  \vek{a}_3 = \tvek{0 \\ 0 \\ c} ,
\end{equation}
with lattice constants $a$ and $c$. The two inequivalent corner
points of the 2D Brillouin zone are then
\begin{equation}
  \label{eq:Kpoint}
  \vek{K}  = \frac{2\pi}{a} \tvek{2/3 \\ 0}, \quad
  \vek{K}' = \frac{2\pi}{a} \tvek{-2/3 \\ 0}.
\end{equation}

We consider a tight-binding Hamiltonian for the BLG $\pi$ bonds
formed by the carbon $p_z$ orbitals, taking into account
nearest-neighbor and second-nearest-neighbor interactions in-plane
and out-of-plane. For a given atom in a honeycomb layer, the vectors
connecting nearest-neighbor atoms are ($j=1,2,3$)
\begin{equation}
  \label{eq:nearest-1}
  \vek{\tau}^{(j)}_1 = \mathcal{R} (2j\pi/3) \, \vek{\tau}^{(3)}_1,\quad
  \vek{\tau}^{(3)}_1 = a \tvek{0 \\ 1/\sqrt{3} \\ 0} ,
\end{equation}
where $\mathcal{R}(\phi)$ denotes a 2D rotation in the $xy$ plane by
the angle $\phi$. Similarly, we get the vectors connecting
second-nearest-neighbor atoms ($j=1,\ldots,6$)
\begin{equation}
  \label{eq:nearest-2}
  \vek{\tau}^{(j)}_2 = \mathcal{R} (j\,\pi/3) \, \vek{a}_1.
\end{equation}
Then the tight-binding Hamiltonian becomes (in the order $A$, $B$,
$A'$, $B'$) \cite{wal47}
\begin{equation}
  \label{eq:tb-ham}
  \begin{array}[b]{@{}l@{}}
  \mathcal{H} (\kvek) = \tilde{E}_0 + \gamma_0' f_2 \\[1ex]
  \hspace{1em} + \tvek[cccc]{
   \Delta 
   & - \gamma_0 \, f_1
   & \gamma_1 
   & \gamma_4  f_1^\ast \\[1.5ex]
     - \gamma_0 \, f_1^\ast
   & 0
   & \gamma_4  f_1^\ast
   &  \gamma_3  f_1 \\[1.5ex]
     \gamma_1 
   &  \gamma_4  f_1
   & \Delta 
   & - \gamma_0 \, f_1^\ast \\[1.5ex]
     \gamma_4  f_1
   & \gamma_3  f_1^\ast
   & - \gamma_0 \, f_1
   & 0 } ,
\end{array}
\end{equation}
where $\tilde{E}_0$ is the site energy of the $p_z$ orbitals,
$\Delta$ denotes the difference between the site energies of $A$
atoms compared with $B$ atoms, $\gamma_0$ ($\gamma_0'$) is the
transfer integral for nearest (second-nearest) neighbors within each
layer, and $\gamma_1$, $\gamma_3$, and $\gamma_4$ are transfer
integrals for atoms in neighboring layers. The functions
$f_l (\kvek)$ are given by
\begin{equation}
  \label{eq:tb-phase}
  f_l (\kvek) \equiv \sum_j e^{i\kvek\cdot\vek{\tau}^{(j)}_l} .
\end{equation}
The particular geometry (\ref{eq:nearest-1}) gives for $f_1 (\kvek)$
\begin{equation}
  \label{eq:tb-fun-1}
  f_1 (\kvek) = e^{i\kcomp_y a / \sqrt{3}}
  + 2e^{-i\kcomp_y a/2\sqrt{3}} \cos (\kcomp_x a /2) ,
\end{equation}
and we have the relation
\begin{equation}
  \label{eq:tb-fun-2}
  f_2 (\kvek) = |f_1 (\kvek)|^2 -3 .
\end{equation}
Thus it is possible to rewrite the Hamiltonian (\ref{eq:tb-ham})
such that it only depends on the function $f_1$. Also, we rearrange
the basis functions in the order $\frac{1}{\sqrt{2}} (A+A')$,
$\frac{1}{\sqrt{2}} (A-A')$, $B$, $B'$ giving

\begin{widetext}
  \begin{equation}
    \label{eq:tb-hamp}
    \mathcal{H} (\kvek) = E_0 + \gamma_0' f_1^2
     + \tvek[cccc]{
     \Delta + \gamma_1  
     & 0
     & \frack{1}{\sqrt{2}} (- \gamma_0 + \gamma_4 ) f_1
     & \frack{1}{\sqrt{2}} (- \gamma_0 + \gamma_4 ) f_1^\ast \\[1.5ex]
     0
     & \Delta - \gamma_1  
     & - \frack{1}{\sqrt{2}} (\gamma_0 + \gamma_4 ) f_1
     &   \frack{1}{\sqrt{2}} (\gamma_0 + \gamma_4 ) f_1^\ast \\[1.5ex]
       \frack{1}{\sqrt{2}} (- \gamma_0 + \gamma_4 ) f_1^\ast
     &  - \frack{1}{\sqrt{2}} (\gamma_0 + \gamma_4 ) f_1^\ast
     & 0
     & \gamma_3  f_1 \\[1.5ex]
       \frack{1}{\sqrt{2}} (- \gamma_0 + \gamma_4 ) f_1
     &  \frack{1}{\sqrt{2}} (\gamma_0 + \gamma_4 ) f_1
     & \gamma_3  f_1^\ast
     & 0 } ,
  \end{equation}
  where $E_0 = \tilde{E}_0 - 3 \gamma_0'$.  In the following, we
  will neglect the constant $E_0$ as well as the small parameter
  $\Delta$.  The latter approximation helps to keep formulas derived
  later on more readable.  Next we expand
  $f_1(\vekc{\kappa}) = f_1 (\vek{K}+\kk)$ around the $\vek{K}$
  point.  Using the coordinate system in Fig.~\ref{fig:lattice} we
  obtain
  \begin{equation}
    f_1 (\kk) = - \frac{\sqrt{3}a}{2} k_-
    + \frac{a^2}{4} k_+^2
    + \ldots \quad .
  \end{equation}
  Substituting this into Eq.\ (\ref{eq:tb-hamp}) gives the
  Slonczewski-Weiss-McClure (SWM) Hamiltonian~\cite{mcc57}
  \begin{equation}
    \label{eq:tb-hams}
    \mathcal{H}^\vek{K}_\kk (\kk) = 
    \tvek[cccc]{
    \gamma_1 + \tgamma_0' k^2
    & 0
    & \frack{1}{\sqrt{2}} (\tgamma_0 - \tgamma_4) k_-
    & \frack{1}{\sqrt{2}} (\tgamma_0 - \tgamma_4) k_+ \\[1.5ex]
    0
    & -  \gamma_1 + \tgamma_0' k^2
    & \frack{1}{\sqrt{2}} (\tgamma_0 +  \tgamma_4) k_-
    & - \frack{1}{\sqrt{2}} (\tgamma_0 +  \tgamma_4) k_+ \\[1.5ex]
        \frack{1}{\sqrt{2}} (\tgamma_0 -  \tgamma_4) k_+
    & \frack{1}{\sqrt{2}} (\tgamma_0 +  \tgamma_4) k_+
    & \tgamma_0' k^2
    & - \tgamma_{31} k_- +  \tgamma_{32} k_+^2 \\[1.5ex]
        \frack{1}{\sqrt{2}} (\tgamma_0 - \tgamma_4) k_-
    & - \frack{1}{\sqrt{2}} (\tgamma_0 + \tgamma_4) k_-
    & - \tgamma_{31} k_+ +  \tgamma_{32} k_-^2
    & \tgamma_0' k^2} .
  \end{equation}
\end{widetext}
Here $(\mathcal{H}^\vek{K}_\kk)_{11}$ describes the uppermost band
at the $\vek{K}$ point transforming according to the irreducible
representation $\Gamma_1$ of $D_3$ (we follow the notation of Koster
\etal~\cite{kos63}), $(\mathcal{H}^\vek{K}_\kk)_{22}$ corresponds to
the lowest band transforming according to $\Gamma_2$ of $D_3$, and
the lower right $2\times 2$ block corresponds to a band transforming
according to $\Gamma_3$ of $D_3$ which is two-fold degenerate at the
$\vek{K}$ point.  We have
\begin{equation}
  \arraycolsep 0.2em
  \begin{array}[b]{rcl*{2}{s{1em}rcl}}
  \tgamma_0 & \equiv & \frack{\sqrt{3}a}{2} \gamma_0, &
  \tgamma_0' & \equiv & \frack{3a^2}{4} \gamma_0',  \\[2ex]
  \tgamma_{31} & \equiv & \frack{\sqrt{3}a}{2} \gamma_3, &
  \tgamma_{32} & \equiv & \frack{a^2}{4} \gamma_3, &
  \tgamma_4    & \equiv & \frack{\sqrt{3}a}{2} \gamma_4.
  \end{array}
\end{equation}
Projecting $\mathcal{H} (\kk)$ on the $\Gamma_3$ subspace using
quasi-degenerate perturbation theory \cite{bir74, win03} gives in
lowest order the effective $2 \times 2$ Hamiltonian [cf.\ Eq.\
(\ref{eq:fieldHam})]
\begin{equation}\label{eq:effM_BLG}
  \begin{array}[b]{r>{\displaystyle}l}
(\mathcal{H}^\vek{K}_\kk)_{33} (\kk) = & 
\frac{\hbar^2}{2 m_0}\left[ - u
\left(k_+^2 \, \sigma_+ + k_-^2 \, \sigma_- \right) + w\, k^2
\sigma_0  \right] \\[1.5ex] &
 - \hbar v \left( k_- \, \sigma_+ + k_+ \, \sigma_- \right)
- \frac{g_\mathrm{m}}{2} \, \mu_\mathrm{B} \, B_z \, \sigma_z \; .
  \end{array}
\end{equation}
[To also obtain the $\Ee_z$-dependent terms in Eq.~(\ref{eq:fieldHam})
that are absent in Eq.\ (\ref{eq:effM_BLG}), a more refined analysis is
required, see Sec.~\ref{sec:loewdin} below.] Defining the energy scale
\begin{subequations}
  \begin{equation}
    \gamma_a = \frac{2\hbar^2}{3 m_0 a^2}
    \approx 0.85~\mathrm{eV} \; ,
  \end{equation}
  we can express the new parameters in terms of the tight-binding
  parameters as follows:
  \begin{align}
    u &= \frac{\gamma_0^2 + \gamma_4^2 -\gamma_1\gamma_3/3}
                            {\gamma_1\gamma_a}
    & & \approx 33 \quad , \\[1ex]
    w &= \frac{\gamma_0' \gamma_1+ 2\gamma_0\gamma_4}
                            {\gamma_1\gamma_a}
    & & \approx 3.4 \quad , \\[1ex]
    v &= \frac{\tgamma_{31}}{\hbar} = \frack{\sqrt{3}a}{2\hbar} \gamma_3
    & & \approx \nume{8.1}{4}{m\, s^{-1}} \quad , \\[1ex] 
    \label{eq:effGfact}
    g_\mathrm{m} &= \frac{4\gamma_0\gamma_4}{\gamma_1 \gamma_a}
    & & \approx 6.2 \quad .
  \end{align}
\end{subequations}
Here we assumed $\gamma_0 = 3.0$~eV, $\gamma_0' = 0.22$~eV,
$\gamma_4 = 0.14$~eV, $\gamma_1 = 0.32$~eV, $\gamma_3 = 0.25$~eV,
and $a = 0.245$~nm.

\section{Symmetry analysis and invariant expansion for BLG}
\label{sec:invExp}

The SWM-like Hamiltonians (\ref{eq:tb-hams}) and (\ref{eq:effM_BLG})
give the band structure of BLG as a function of the wave vector
$\kk$ measured from the $\vek{K}$ point. The theory of invariants
\cite{bir74, win10a} allows one to generalize these results to
perturbations $\vekc{K}$ that involve combinations of various
quantities in addition to the wave vector $\kk$, e.g., electric and
magnetic fields $\vekc{E}$ and $\vek{B}$, strain $\Strain$ and the
intrinsic spin $\vek{s}$.  The group of the wave vector $\vek{K}$ is
isomorphic to the trigonal point group $D_3$ so that we can classify
the electronic states near $\vek{K}$ using the irreducible
representations of $D_3$ \cite{kos63}. The $4\times 4$ Hamiltonian
$\mathcal{H}^\vek{K}$ falls into blocks
\begin{equation}
  \label{eq:swm:blocks}
  \mathcal{H}^\vek{K} = \tvek[ccc]{
      \mathcal{H}_{11} & \mathcal{H}_{12} & \mathcal{H}_{13} \\
      \mathcal{H}_{21} & \mathcal{H}_{22} & \mathcal{H}_{23} \\
      \mathcal{H}_{31} & \mathcal{H}_{32} & \mathcal{H}_{33}} ,
\end{equation}
where each diagonal block $\mathcal{H}_{\alpha\alpha}$ describes a
band transforming according to the irreducible representation
$\Gamma_\alpha$ of $D_3$.  In the end, we are mainly interested in
the block $\mathcal{H}_{33}$ corresponding to the highest valence
band and lowest conduction band of BLG.  However, for the study of
prefactors given in Sec.~\ref{sec:loewdin}, all bands in the SWM
model are included.

According to the theory of invariants, each block
$\mathcal{H}_{\alpha\beta}$ takes the form
\begin{equation}
  \label{eq:invar}
  \mathcal{H}_{\alpha\beta} (\vekc{K})
  =  \sum_{\kappa, \, \lambda}
  \koeff{a}{\alpha\beta}{\kappa\lambda}
  \sum_{l=1}^{L_\kappa} {X}_l^{(\kappa,\alpha\beta)}
  \mathcal{K}_l^{(\kappa,\lambda) \, \ast} \, .
\end{equation}
Here $\koeff{a}{\alpha\beta}{\kappa\lambda}$ are prefactors,
${X}_l^{(\kappa,\alpha\beta)}$ are matrices that transform according
to the IRs $\Gamma_\kappa$ (of dimension $L_\kappa$) contained in
the product representation $\Gamma_\alpha \times \Gamma_\beta^\ast$
of $D_3$. Likewise, $\vekc{K}$ can be decomposed into irreducible
tensor operators $\vekc{K}^{(\kappa,\lambda)}$ that transform
according to the IRs $\Gamma_{\kappa}$ of $D_3$. Using the
coordinate system in Fig.~\ref{fig:lattice} we obtain the basis
matrices and tensor operators listed in Tables~\ref{tab:basemat} and
\ref{tab:tensorop}.  For completeness, Table~\ref{tab:tensorop} also
includes the lowest-order tensor operators due to strain.  However,
in the following, these strain tensor operators are not considered
further.  Quite generally, each term proportional to the components
of the strain tensor $\strain_{ij}$ is formally equivalent to a term
where $\strain_{ij}$ is replaced by the symmetrized product
$\{k_i,k_j\}$ \cite{bir74}.


\begin{table}
  \caption[]{\label{tab:basemat} Symmetrized matrices for the
  invariant expansion of the blocks $\mathcal{H}_{\alpha\beta}$ for
  the point group $D_{3}$.}
$\arraycolsep 0.8em
\renewcommand{\arraystretch}{1.2}
\begin{array}{clcl} \hline \hline
\mbox{Block} &
\multicolumn{1}{l}{\mbox{Representations}} &
\multicolumn{2}{c}{\mbox{Symmetrized matrices}} \\ \hline
\mathcal{H}_{11} & \Gamma_1 \times \Gamma_1^\ast = \Gamma_1 &
    \Gamma_1: & (1)  \\
\mathcal{H}_{22} & \Gamma_2 \times \Gamma_2^\ast = \Gamma_1 &
    \Gamma_1: & (1)  \\
\mathcal{H}_{12} & \Gamma_1 \times \Gamma_2^\ast = \Gamma_2 &
    \Gamma_2: & (1)  \\
\mathcal{H}_{13} & \Gamma_1 \times \Gamma_3^\ast = \Gamma_3 &
    \Gamma_3: & (1,1), (-i,i)  \\
\mathcal{H}_{23} & \Gamma_2 \times \Gamma_3^\ast = \Gamma_3 &
    \Gamma_3: & (1,-1), (-i,-i)  \\
\mathcal{H}_{33} & \Gamma_3 \times \Gamma_3^\ast &
    \Gamma_1: & \openone  \\
& = \Gamma_1 + \Gamma_2 + \Gamma_3
  & \Gamma_2: & \sigma_z \\
  & & \Gamma_3: & \sigma_x, \sigma_y
\\ \hline \hline
\end{array}$
\end{table}

\begin{table}
  \caption[]{\label{tab:tensorop} Irreducible tensor components for
  the point group $D_{3}$ (the group of the $\vek{K}$ point in
  BLG). Terms printed in bold give rise to invariants in the block
  $\mathcal{H}^\vek{K}_{33}$ allowed by time-reversal
  invariance. (No terms proportional to $k_z$ are listed as they are
  irrelevant for graphene.)  Contributions that are new in BLG
  (i.e., are not part of the corresponding set~\cite{win10a} for
  $D_{3h}$ which is the group of the $\vek{K}$ point in SLG) are
  marked by $\ddagger$.  Notation:
  $\{A, B\} \equiv \frac{1}{2} (AB + BA)$.}
  $\extrarowheight 0.2ex
  \newcommand{\nl}{\newline}
  \begin{array}{cs{1.0em}p{0.9\columnwidth}} \hline \hline
    \Gamma_1 & \allowed{1};
    \allowed{k_x^2 + k_y^2};
    \allowed{\{k_x, 3k_y^2 - k_x^2\}};
    \forbidden{B_x k_x + B_y k_y}; \nl
    \forbidden{k_x \Ee_x + k_y \Ee_y};
    \allowednew{\Ee_x B_x + \Ee_y B_y};
    \allowednew{\Ee_z B_z};
    \allowed{\strain_{xx} + \strain_{yy}};
    \allowednew{\strain_{zz}}; \nl
    \allowed{(\strain_{yy} - \strain_{xx}) k_x + 2 \strain_{xy} k_y};
    \allowednew{\strain_{yz} k_x - \strain_{xz} k_y}; \nl
    \forbidden{(\strain_{yy} - \strain_{xx}) B_x + 2 \strain_{xy} B_y};
    \forbidden{\strain_{yz} B_x - \strain_{xz} B_y}; \nl
    \forbidden{(\strain_{yy} - \strain_{xx}) \Ee_x + 2 \strain_{xy} \Ee_y};
    \forbidden{\strain_{yz} \Ee_x - \strain_{xz} \Ee_y};
    \forbidden{s_x k_x + s_y k_y}; \nl
    \allowed{s_x B_x + s_y B_y};
    \allowed{s_z B_z};
    \allowednew{s_x \Ee_x + s_y \Ee_y};
    \allowednew{s_z \Ee_z}; \nl
    \allowed{(s_x k_y - s_y k_x) \Ee_z};
    \allowed{s_z (k_x \Ee_y - k_y \Ee_x)}; \nl
    \forbidden{s_x (\strain_{yy} - \strain_{xx}) + 2 s_y \strain_{xy}};
    \forbidden{s_x \strain_{yz} - s_y \strain_{xz}} \\
    \Gamma_2 & \forbidden{\{k_y, 3k_x^2-k_y^2\}};
    \allowed{B_z};
    \allowednew{k_x B_y - k_y B_x};
    \allowednew{\Ee_z};
    \allowed{k_x \Ee_y - k_y \Ee_x}; \nl
    \forbidden{\Ee_x B_y - \Ee_y B_x};
    \forbidden{(\strain_{xx} - \strain_{yy}) k_y + 2 \strain_{xy} k_x};
    \forbidden{\strain_{yz} k_y + \strain_{xz} k_x}; \nl
    \allowednew{(\strain_{xx} - \strain_{yy}) B_y + 2 \strain_{xy} B_x};
    \allowed{(\strain_{xx} + \strain_{yy}) B_z};
    \allowednew{\strain_{zz} B_z}; \nl
    \allowednew{\strain_{xz} B_x + \strain_{yz} B_y};
    \allowed{(\strain_{xx} - \strain_{yy}) \Ee_y + 2 \strain_{xy} \Ee_x}; \nl
    \allowednew{(\strain_{xx} + \strain_{yy}) \Ee_z};
    \allowednew{\strain_{zz} \Ee_z};
    \allowednew{\strain_{xz} \Ee_x + \strain_{yz} \Ee_y};
    \allowed{s_z}; \nl
    \allowednew{s_x k_y - s_y k_x};
    \forbidden{s_x B_y - s_y B_x};
    \forbidden{s_x \Ee_y - s_y \Ee_x}; \nl
    \forbidden{(s_x k_x + s_y k_y) \Ee_z};
    \allowednew{s_y (\strain_{xx} - \strain_{yy}) + 2 s_x \strain_{xy}};\nl
    \allowed{s_z (\strain_{xx} + \strain_{yy})};
    \allowednew{s_x \strain_{xz} + s_y \strain_{yz}};
    \allowednew{s_z \strain_{zz}}; \\
    %
    \Gamma_3 &
    \allowed{k_x, k_y};
    \allowed{\{k_y+k_x, k_y - k_x\}, 2 \{k_x, k_y\}}; \nl
    \allowed{\{k_x, k_x^2 + k_y^2\}, \{k_y, k_x^2 + k_y^2\}}; \nl
    \forbidden{B_x, B_y};
    \forbidden{B_y k_y - B_x k_x, B_x k_y + B_y k_x};
    \forbidden{B_z k_y, - B_z k_x}; \nl
    \forbidden{\Ee_x, \Ee_y};
    \forbidden{\Ee_y k_y - \Ee_x k_x, \Ee_x k_y +  \Ee_y k_x};
    \forbidden{\Ee_z k_y, - \Ee_z k_x}; \nl
    \allowednew{\Ee_y B_y - \Ee_x B_x, \Ee_y B_x + \Ee_x B_y};
    \allowed{\Ee_y B_z, - \Ee_x B_z}; \nl
    \allowed{\Ee_z B_y, - \Ee_z B_x};
    \allowed{\strain_{yy} - \strain_{xx}, 2 \strain_{xy}};
    \allowednew{\strain_{yz}, - \strain_{xz}}; \nl
    \allowed{(\strain_{xx} + \strain_{yy}) (k_x, k_y)};
    \allowednew{\strain_{yz} k_x + \strain_{xz} k_y,
     \strain_{xz} k_x  - \strain_{yz} k_y}; \nl
    \allowed{(\strain_{xx} - \strain_{yy}) k_x + 2\strain_{xy} k_y,
     (\strain_{yy} - \strain_{xx}) k_y + 2\strain_{xy} k_x}; \nl
    \allowednew{\strain_{zz} k_x, \strain_{zz} k_y};
    \nl
    \forbidden{(\strain_{xx} + \strain_{yy}) (B_x, B_y)};
    \forbidden{\strain_{yz} B_x + \strain_{xz} B_y,
     \strain_{xz} B_x  - \strain_{yz} B_y};  \nl
    \forbidden{(\strain_{xx} - \strain_{yy}) B_x + 2 \strain_{xy} B_y,
     (\strain_{yy} - \strain_{xx}) B_y + 2 \strain_{xy} B_x}; \nl
    \forbidden{2 \strain_{xy} B_z, (\strain_{xx} - \strain_{yy}) B_z};
    \forbidden{\strain_{zz} B_x, \strain_{zz} B_y};
    \forbidden{\strain_{xz} B_z, \strain_{yz} B_z}; \nl
    \forbidden{(\strain_{xx} + \strain_{yy}) (\Ee_x, \Ee_y)};
    \forbidden{\strain_{yz} \Ee_x + \strain_{xz} \Ee_y,
     \strain_{xz} \Ee_x  - \strain_{yz} \Ee_y}; \nl
    \forbidden{(\strain_{xx} - \strain_{yy}) \Ee_x + 2 \strain_{xy} \Ee_y,
     (\strain_{yy} - \strain_{xx}) \Ee_y + 2 \strain_{xy} \Ee_x}; \nl
    \forbidden{2 \strain_{xy} \Ee_z, (\strain_{xx} - \strain_{yy}) \Ee_z};
    \forbidden{\strain_{zz} \Ee_x, \strain_{zz} \Ee_y};
    \forbidden{\strain_{xz} \Ee_z, \strain_{yz} \Ee_z}; \nl
    \forbidden{s_x, s_y};
    \forbidden{s_y k_y - s_x k_x, s_x k_y + s_y k_x};
    \forbidden{s_z k_y, - s_z k_x}; \nl
    \allowed{s_y B_y - s_x B_x, s_x B_y + s_y B_x};
    \allowednew{s_z B_y, - s_z B_x}; \nl
    \allowednew{s_y B_z, - s_x B_z};
    \allowednew{s_y \Ee_y - s_x \Ee_x, s_x \Ee_y + s_y \Ee_x}; \nl
    \allowed{s_z \Ee_y, - s_z \Ee_x};
    \allowed{s_y \Ee_z, - s_x \Ee_z}; \nl
    \allowed{s_z (k_x \Ee_y + k_y \Ee_x), s_z (k_x \Ee_x - k_y \Ee_y)}; \nl
    \allowednew{(k_x, k_y) s_z \Ee_z}; 
    \allowed{(s_x k_y + s_y k_x, s_x k_x - s_y k_y) \Ee_z}; \nl
    \forbidden{(s_x, s_y) (\strain_{xx} + \strain_{yy})};
    \forbidden{2 s_z \strain_{xy}, s_z (\strain_{xx} - \strain_{yy})}; \nl
    \forbidden{s_x (\strain_{xx} - \strain_{yy}) - 2 s_y \strain_{xy},
     s_y (\strain_{yy} - \strain_{xx}) - 2 s_x \strain_{xy}}; \nl
    \forbidden{s_x \strain_{zz}, s_y \strain_{zz}};
    \forbidden{s_z \strain_{xz}, s_z \strain_{yz}};
    \forbidden{s_x \strain_{yz} + s_y \strain_{xz},
     s_x \strain_{xz}  - s_y \strain_{yz}};
    \\ \hline \hline
  \end{array}$
\end{table}


Additional constraints for the Hamiltonian (\ref{eq:swm:blocks}) are
due to time reversal invariance. The crystallographic point group
$D_{3d}$ of BLG contains symmetry elements $R$ mapping the basis
functions $\Psi_{\vek{K} \lambda}$ at $\vek{K}$ on
$\Psi_{\vek{K}' \lambda}$ at $\vek{K}'$. These basis functions are
also mapped onto each other by time reversal $\theta$, i.e., we have
\begin{equation}
  \label{eq:time}
  \theta \, \Psi_{\vek{K},\lambda} = \Psi_{\vek{K}\lambda}^\ast
  = \sum_{\lambda'} \mathcal{T}_{\lambda\lambda'}
  \: \Psi_{\vek{K}'\lambda'} \,,
\end{equation}
with a unitary matrix~$\mathcal{T}$. Combining these operations we
obtain \cite{bir74, win10a, man07}
\begin{equation}
  \label{eq:timea2}
  \mathcal{T}^{-1} \mathcal{H} (R^{-1} \vekc{K}) \mathcal{T}
  = \mathcal{H}^\ast (\zeta \vekc{K}) = \mathcal{H}^t (\zeta \vekc{K}) \,,
\end{equation}
where $\ast$ denotes complex conjugation and $t$ is transposition.
The quantity $\zeta$ depends on the behavior of $\vekc{K}$ under
time reversal.  The vectors $\kk$, $\vek{B}$, and $\vek{s}$ are odd
under time reversal so that then $\zeta=-1$, while $\vekc{E}$ and
$\Strain$ have $\zeta=+1$. Equation (\ref{eq:timea2}) provides a
general criterion for determining which terms in the expansion
(\ref{eq:invar}) are allowed by time-reversal invariance and which
terms are forbidden. For off-diagonal blocks
$\mathcal{H}_{\alpha\beta}$, the criterion (\ref{eq:timea2}) also
determines the phase of the respective prefactors
$\koeff{a}{\alpha\beta}{\kappa\lambda}$.  The matrix~$\mathcal{T}$
depends on the choice for the operation $R$. If $R$ is the
reflection $R_y$ at the $yz$ plane [thus mapping the atoms in each
sublattice in each layer onto each other, see
Fig.~\ref{fig:lattice}(a)], the matrix~$\mathcal{T}$ is simply the
identity matrix and we obtain
\begin{equation}
  \label{eq:reflect}
  \mathcal{H}^\vek{K} (R_y^{-1}\vekc{K})
  = \mathcal{H}^{\vek{K}\,\ast} (\zeta \vekc{K}) \quad .
\end{equation}
Those tensor operators in Table~\ref{tab:tensorop} that satisfy the
criterion (\ref{eq:reflect}) for the block
$\mathcal{H}^\vek{K}_{33}$ are printed in bold face.

We note that under $R_y$ polar ($\vek{p}$) and axial ($\vek{a}$)
vectors transform as
\begin{subequations}
  \label{eq:transVal}
  \begin{align}
    \label{eq:transPol}
    p_x & \to -p_x \quad , & \quad p_{y, z} & \to p_{y, z} \, \quad ,\\
    \label{eq:transAxi}
    a_x & \to a_x \quad , & \quad a_{y, z} & \to -a _{y, z} \, \quad .
  \end{align}
\end{subequations}
The transformational properties for the components of the
second-rank strain tensor $\strain_{ij}$ can be expressed similarly.
To illustrate valley-dependent physics, we will often employ a
compact notation where $\tau_0$ ($\tau_z$) denotes the unit
(diagonal Pauli) matrix acting in valley-isospin space. The rules in
Eq.~(\ref{eq:transVal}) can then be expressed by writing general
vector operators as $(p_x\,\tau_z, p_y\, \tau_0, p_z\,\tau_0)$ and
$(a_x\, \tau_0, a_y\, \tau_z, a_z\, \tau_z)$.

The group $D_3$ characterizing the $\vek{K}$ point in BLG is a
subgroup of the group $D_{3h}$ for the $\vek{K}$ point in SLG so
that any term allowed by spatial symmetries in
$\mathcal{H} (\vekc{K})$ for SLG is likewise allowed in
BLG. Moreover, the constraint (\ref{eq:timea2}) due to time reversal
invariance is exactly equivalent to the constraint in SLG. Thus it
follows immediately that the invariant expansion for the $\Gamma_3$
band of BLG contains all terms that exist already for SLG (though
the respective prefactors are unrelated).  In particular, this
yields immediately the Hamiltonian (\ref{eq:effM_BLG}), which
applies also to SLG, the only difference being that for SLG the
$k$-linear term proportional to $\hbar v$ is dominant, whereas the
$k$-quadratic terms are small.  In BLG, the situation is reversed:
for typical Fermi wave vectors the dispersion is dominated by the
quadratic terms, whereas the $k$-linear term is only a small
correction.

\section{Magneto-Electric Equivalence}
\label{sec:MEanal}

A more detailed analysis shows that the point group $D_{3h}$ for the
$\vek{K}$ point of SLG distinguishes, as is usual, between polar
vectors (such as the electric field $\vekc{E}$) and axial vectors
(such as the magnetic field $\vek{B}$). Thus each term in the
$2\times 2$ SLG Hamiltonian with a certain functional form and
linear in the field $\vekc{E}$ or $\vek{B}$ is forbidden for the
other field. However, in BLG the point group $D_3$ does not
distinguish between polar and axial vectors, the reason being that
$D_3$ only contains rotations as symmetry elements. The $x$ and $y$
components of \emph{any} vector transform according to $\Gamma_3$,
whereas the $z$ component transforms according to $\Gamma_2$. This
implies that spatial symmetries cannot distinguish electric and
magnetic fields in BLG. Moreover, Eq.\ (\ref{eq:timea2}) treats
electric and magnetic fields symmetrically, too. Thus it follows
that every $\vekc{E}$-dependent term in the BLG Hamiltonian
(\ref{eq:invar}) is accompanied by another term where $\vekc{E}$ is
simply replaced by $\vek{B}$ (and vice versa for $\vek{B}$-dependent
terms). However, the prefactors of these terms are, in general,
unrelated (see Sec.~\ref{sec:loewdin}).  Table~\ref{tab:invariants}
summarizes the new terms arising in lowest order from this
magneto-electric equivalence.  The well-known fact that a potential
difference between the layers induces a gap~\cite{mcc06, mcc06b,
oht06, zha09} is embodied by the term $\propto\Ee_z \sigma_z$, which
is the electric analog of the orbital Zeeman splitting. Moreover, we
obtain a rather counter-intuitive purely electric-field-dependent
spin splitting~\cite{win12, gei13, gmi13} $\propto\Ee_z s_z$. Thus
real spins and pseudospins in BLG can precess not only in a magnetic
field but also in an electric field.  In second order of the fields,
we also get terms proportional to $\Ee_z B_z$ [see Eq.\
(\ref{eq:fieldHam})] and $\vekc{E}_\| \cdot \vek{B}_\|$ reminiscent
of the electrodynamics in axion field theory~\cite{wil87}.  The
magnetoelectric effect arising in BLG from these second-order terms
has been discussed in Ref.~\onlinecite{zue14}.

\begin{table*}[t]
  \caption{\label{tab:invariants} Lowest-order terms (excluding
  strain-induced couplings) reflecting magneto-electric equivalence
  in BLG. The upper sign holds for valley $\vek{K}$, the lower sign
  for $\vek{K}'$. The different behavior of electric and magnetic
  fields under inversion at the $yz$ plane results in the opposite
  valley symmetry of magneto-electric analogs, which are therefore
  most generally related by the combined replacements
  $\vek{\Ee}\leftrightarrow\vek{B}$ and
  $\tau_0\leftrightarrow\tau_z$.  Terms present in BLG but not in
  SLG are marked by $\ddagger\,$.}
  \renewcommand{\arraystretch}{1.3}
  \begin{tabular}{rRs{0.6em}cs{0.6em}Ll} \hline \hline
\multicolumn{2}{c}{magnetic field $\vek{B}$} &
& \multicolumn{2}{c}{electric field $\vekc{E}$} \\ \hline
orbital Zeeman splitting &
B_z\,\sigma_z\,\tau_z &
(T1) &
\Ee_z \,\sigma_z\,\tau_0 &
inter-layer (pseudospin) gap$\ddagger$ \\
orbital Zeeman splitting$\ddagger$ &
(i/2)(k_+ B_- - k_- B_+) \sigma_z \,\tau_0 &
(T2) & (i/2)(k_+ \Ee_- - k_- \Ee_+) \sigma_z \,\tau_z &
orbital Rashba splitting \\
spin Zeeman splitting &
B_z\, s_z\,\tau_0 &
(T3) &
\Ee_z \, s_z\, \tau_z &
electric spin splitting$\ddagger$ \\
spin-orbital Zeeman splitting$\ddagger$ &
2i B_z (s_- \sigma_\pm - s_+ \sigma_\mp)\,\tau_0 &
(T4) &
2i \Ee_z (s_- \sigma_\pm - s_+ \sigma_\mp)\,\tau_z &
Rashba spin splitting \\
spin-orbital Zeeman splitting$\ddagger$ &
i B_z (s_+ k_- - s_- k_+) \,\tau_z &
(T5) &
i \Ee_z (s_+ k_- - s_- k_+) \,\tau_0 &
Rashba spin splitting \\
spin-orbital Zeeman splitting$\ddagger$ &
2i B_z (s_- k_- \sigma_\mp - s_+ k_+ \sigma_\pm)\,\tau_z &
(T6) &
2i \Ee_z (s_- k_- \sigma_\mp - s_+ k_+ \sigma_\pm) \,\tau_0 &
Rashba spin splitting \\
spin Zeeman splitting &
(B_+ s_- + B_- s_+) \, \tau_0  &
(T7) &
(\Ee_+ s_- + \Ee_- s_+) \, \tau_z &
electric spin splitting$\ddagger$ \\
spin-orbital Zeeman splitting$\ddagger$ &
i (B_- \sigma_\pm - B_+ \sigma_\mp) s_z \, \tau_0 &
(T8) &
i (\Ee_- \sigma_\pm - \Ee_+ \sigma_\mp) s_z \, \tau_z &
Rashba spin splitting \\
spin-orbital Zeeman splitting &
-2 (B_+ s_+ \sigma_\pm + B_- s_- \sigma_\mp) \, \tau_0 &
(T9) &
-2 (\Ee_+ s_+ \sigma_\pm + \Ee_- s_- \sigma_\mp) \, \tau_z &
Rashba spin splitting$\ddagger$ 
\\ \hline \hline
\end{tabular}
\end{table*}

We note that the Hamiltonian $\mathcal{H}^\vek{K} (\vekc{K})$
depends not only on the fields $\vek{B}$ and $\vekc{E}$ but also on
the electrodynamic potentials $\vek{A}$ and $\Phi$. The latter terms
are not affected by the magneto-electric equivalence.

\subsection{Valley Dependence}
\label{sec:valley}

The intravalley dynamics induced by the $\vek{B}$ and $\vekc{E}$
dependent terms in Table~\ref{tab:invariants} are indistinguishable
on a qualitative level. They differ only in the magnitude of the
induced effects. However, differences do arise when comparing the
dynamics in the two valleys $\vek{K}$ and $\vek{K}'$.  The effective
Hamiltonian for electrons in the $\vek{K}'$ valley can be obtained
from that for the $\vek{K}$ valley by a reflection $R_y$ of the
vectors $\kk$, $\vek{s}$, $\vekc{E}$, and $\vek{B}$ at the $yz$
plane, see Fig.~\ref{fig:lattice}(a) and Eq.\ (\ref{eq:transVal})
\cite{win10a}.  Choosing the convention that the $\vek{B}$ and
$\vekc{E}$-dependent terms have the same sign in the $\vek{K}$
valley, the corresponding term in the $\vek{K}'$ valley involving
the axial vector $\vek{B}$ (second column in
Table~\ref{tab:invariants}) differs by an overall minus sign from
the term involving the polar vector $\vekc{E}$ (fourth column in
Table~\ref{tab:invariants}).  For the valley dependence resulting
for the terms in Eq.\ (\ref{eq:eff_BLG}), see
Table~\ref{tab:prefact} below.

The Hamiltonian for valley $\vek{K}'$ can be obtained from
$(\mathcal{H}^\vek{K}_\kk)_{33} (\kk)$ using the transformation
described above.  Alternatively, considering all possible
interactions for either a magnetic field $\vek{B}$ or an electric
field $\vekc{E}$, the Hamiltonian for one valley can be derived from
the Hamiltonian for the other valley via a simple (anti-) unitary
transformation.  Using our phase conventions, the relation
\begin{subequations}
\label{eq:valtrafo}
\begin{equation}
  (\mathcal{H}^{\vek{K}'}_\kk)_{33} (\kk) = 
  \sigma_x \, (\mathcal{H}^\vek{K}_\kk)_{33} (-\kk) \, \sigma_x
\end{equation}
holds if only $\vek{B} \ne 0$.  In contrast, with solely an electric
field $\vekc{E}$ present, we have
\begin{equation}
  (\mathcal{H}^{\vek{K}'}_\kk)_{33} (\kk) = 
  s_y \, (\mathcal{H}^\vek{K}_\kk)_{33}^\ast (-\kk) \, s_y \;,
\end{equation}
\end{subequations}
where $\ast$ denotes complex conjugation.  The relations
(\ref{eq:valtrafo}) imply that with all matter-field interactions
associated with just one field taken into account, the valley
degeneracy is preserved.  However, the valley degeneracy will
generally be broken when $\vek{B}$ and $\vekc{E}$ fields are applied
simultaneously \cite{xia07}.

\subsection{Thermodynamic Response}

The measurable macroscopic response of a material to an externally
applied magnetic (electric) field is characterized by the
magnetization density $\vekc{M}$ (dielectric polarization density
$\vekc{P}$).  Using $\vekc{F}$ to denote either $\vek{B}$ or
$\vekc{E}$, and introducing $\vekc{R}$ to be the associated response
$\vekc{M}$ or $\vekc{P}$, we have at temperature $T=0$
(Ref.~\onlinecite{ash76})
\begin{equation}\label{eq:response}
  \vekc{R} (\vekc{F}) = - \frac{1}{V} \frac{\partial E_0
  (\vekc{F})}{\partial \vekc{F}} \quad .
\end{equation}
Here $V$ is the system volume and $E_0 (\vekc{F})$ is the
many-particle ground state energy of the system as a function of the
field $\vekc{F}$.  Quite generally, the thermodynamic response can
be probed experimentally, e.g., by a spatially inhomogenous field
$\vekc{F} (\rr)$ that varies slowly over the sample.  The vector
gradient $\vek{\nabla} \vekc{F} (\rr)$ (a second-rank tensor)
gives rise to a force per unit volume $\vek{f}$ exerted on the
system \cite{ash76}
\begin{equation}
  \vek{f} = - \frac{\nabla E_0 [ \vekc{F} (\rr)]}{V}
  = \left[\vek{\nabla} \vekc{F} (\rr) \right]
    \cdot \vekc{R} (\vekc{F}) \quad .
\end{equation}
This can be viewed as a generalized Stern-Gerlach experiment.

According to Eqs. (\ref{eq:valtrafo}), when applying either a
magnetic field $\vek{B}$ or electric field $\vekc{E}$, the
functional form of the field-induced changes in the energy spectrum
are the same in both valleys for both $\vek{B}$ and $\vekc{E}$.
Accordingly, the response functions (\ref{eq:response}) are also the
same (apart from numeric prefactors) when considering each term in
Table \ref{tab:invariants} for magnetic and electric fields.  For
example, the Zeeman spin splitting described by the magnetic term
(T3) results in the well-known paramagnetic Pauli contribution
\cite{ash76} to the magnetization density of a conductor at
temperatures $T\ll E_\mathrm{F}/k_\mathrm{B}$,
\begin{subequations}
  \label{eq:mag-pol}
  \begin{equation}
    \label{eq:magnetization}
    \mathcal{M}_z^{(\mathrm{T3})} =
    \left(\frac{g_\mathrm{m}^{(\mathrm{s})} \mu_\mathrm{B}}{2}\right)^2
    \, B_z \, D (E_F) \quad ,
  \end{equation}
  where $D (E_F)$ is the density of states at the Fermi energy $E_F$
  and $g_\mathrm{m}^{(\mathrm{s})}$ the real-spin \textit{g} factor
  for electrons in BLG.  When neglecting for clarity the small
  $k$-linear term proportional to $v$ in the Hamiltonian
  (\ref{eq:fieldHam}), we have
  $D(E_F) = 2 m_0 / [\pi \hbar^2(u \pm w)]$ for electron and hole
  states. The corresponding electric spin splitting (T3) due to a
  perpendicular electric field $\Ee_z$ gives rise to the dielectric
  polarization density
  \begin{equation}
    \label{eq:polarization}
    \mathcal{P}_z^{(\mathrm{T3})} =
    \left(\frac{g_\mathrm{e}^{(\mathrm{s})} \mu_\mathrm{B}}{2}\right)^2 \,
    \frac{\Ee_z}{c^2} \, D (E_F) \quad ,
  \end{equation}
\end{subequations}
in complete analogy with Eq.\ (\ref{eq:magnetization}).  Such an
unusual spin-dependent contribution to the dielectric polarization
density is clearly a consequence of spin-orbit coupling, which is
responsible for a nonzero prefactor $g_\mathrm{e}^{(\mathrm{s})}$.

Writing Eq.~(\ref{eq:response}) more formally in terms of the
many-particle Hamiltonian $H$,
\begin{equation}
  \vekc{R} (\vekc{F}) = - \frac{1}{\beta V}\, 
  \frac{\partial}{\partial \vekc{F}}
  \, \ln \left[ \trace \left\{\exp\left( - \beta\, H
      \right)\right\}\right] \,\, ,
\end{equation}
where $\trace \{\dots\}$ denotes the trace in the many-particle
Hilbert space and $\beta \equiv 1/(k_\mathrm{B} T)$ with Boltzmann
constant $k_\mathrm{B}$, we see that each term in the
single-particle Hamiltonian that is linear in the field $\vekc{F}$
results in a contribution to the response $\vekc{R}$ that
corresponds to the expectation value of the quantity the field
couples to.  For example, the spin contribution to the magnetization
density arising from term (T3) can be written as
\begin{subequations}
\label{eq:parMagx}
\begin{eqnarray}
\mathcal{M}_z^{(\mathrm{T3})} &=& \frac{g_\mathrm{m}^{(\mathrm{s})}
\mu_\mathrm{B}}{V}\, \frac{\trace \left\{  \sum_j s_{zj}\, \tau_{0j}\,
\exp\left( - \beta\, H \right)\right\}}{\trace \left\{\exp\left( - \beta \, H
\right)\right\}} \, , \hspace*{1em} \\
&\equiv& \frac{N}{V} \, g_\mathrm{m}^{(\mathrm{s})} \mu_\mathrm{B}
\left\langle s_z \,\tau_0\right\rangle \, , \label{eq:parMag}
\end{eqnarray}
\end{subequations}
where $\sum_j$ is the sum over all particles and $N$ the particle
number.  Equation~(\ref{eq:parMag}) embodies the conventional
understanding that the spin contribution to the magnetization is
proportional to the thermal average of the microscopic spin-magnetic
moment.  Due to the magneto-electric equivalence in BLG, additional
contributions exist that are associated with unconventional
averages, e.g.,
\begin{subequations}\label{eq:unconvM}
\begin{eqnarray}
\mathcal{M}_z^{(\mathrm{T4})} &\propto& \left\langle i (s_- \sigma_\pm -
s_+ \sigma_\mp)\,\tau_0 \right\rangle \,\, , \\
\mathcal{M}_z^{(\mathrm{T5})} &\propto& \left\langle i (s_+ k_- - s_- k_+)
\,\tau_z \right\rangle \,\, , \\
\mathcal{M}_z^{(\mathrm{T6})} &\propto& \left\langle i (s_- k_- \sigma_- 
- s_+ k_+ \sigma_+)\,\tau_z \right\rangle \,\, .
\end{eqnarray}
\end{subequations}
In order to obtain the contributions to the dielectric polarization
density $\mathcal{P}_z$, we must swap $\tau_0$ and $\tau_z$ in
Eqs.~(\ref{eq:parMagx}) and (\ref{eq:unconvM}).  At the same time,
we also need to use (anti-) unitarily transformed states to evaluate
the thermal averages in the two valleys [see Eq.\
(\ref{eq:valtrafo})] so that, in the end, we get equivalent
expressions for $\mathcal{M}_z$ and $\mathcal{P}_z$.

\subsection{Spin Textures and Spin Polarizations}

\begin{figure*}
  \includegraphics[height=0.86\textheight]{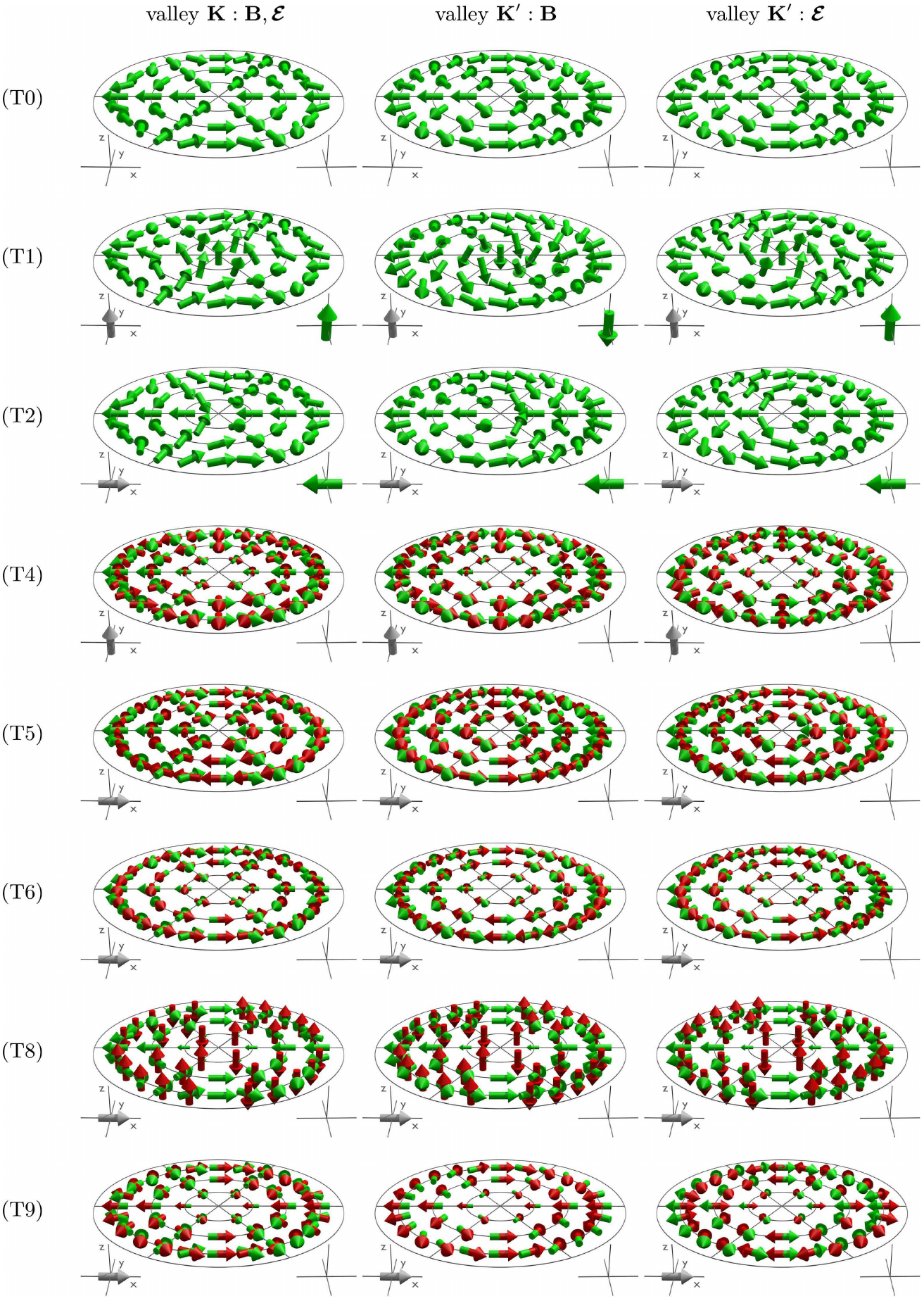}
  \caption{\label{fig:visual}Visualization of real and pseudospin
  textures in BLG generated by field-dependent interactions shown in
  Table~\ref{tab:invariants} [omitting the trivial interactions (T3)
  and (T7)].  The first row denoted (T0) shows the pseudospin
  texture without any external fields.  For selected points in
  $\vek{k}$ space, a red (green) arrow indicates the expectation
  value of the real (pseudo-)spin vector of the corresponding
  negative-energy eigenstate obtained by diagonalizing the leading
  field-independent contribution to the BLG Hamiltonian together
  with the respective term in Table~\ref{tab:invariants}. The 1st
  (2nd, 3rd) column shows results for the valley $\vek{K}$
  ($\vek{K'}$ with $\vek{B}$, $\vek{K'}$ with $\vekc{E}$). Note that
  our sign convention is such that $\vek{B}$ and $\vekc{E}$ have the
  same effect in the $\vek{K}$ valley (1st column). Gray arrows
  indicate the direction of the applied field. Big arrows in the
  lower right of the panels indicate the existence of a net spin
  polarization obtained by averaging over the (pseudo-)spin
  expectation values of occupied hole states for a negative chemical
  potential.}
\end{figure*}

While the energy spectra obtained by either a field $\vek{B}$ or
$\vekc{E}$ are the same in both valleys, Eqs.\ (\ref{eq:valtrafo})
imply that the (pseudo-) spin textures and (pseudo-) spin
polarizations induced by fields $\vek{B}$ and $\vekc{E}$ due to each
term in Table~\ref{tab:invariants} are qualitatively different from
each other, see Fig.~\ref{fig:visual}.  For example, the orbital
Zeeman term $\propto B_z \sigma_z\tau_z$ [i.e., the magnetic term
(T1) in Table~\ref{tab:invariants}] couples the field component
$B_z$ to the pseudospin associated with the sublattice degree of
freedom $\sigma_z$ in SLG and BLG. However, the field $B_z$ does not
induce a global pseudospin (sublattice) polarization because this
term has opposite signs in the two valleys so that the pseudospin
polarization in the two valleys is antiparallel. As a result (and as
to be expected), the two sublattices remain indistinguishable even
for finite $B_z$. Only an electric field $\Ee_z$ polarizes the
sublattices in BLG via the term $\propto \Ee_z\sigma_z\tau_0$,
consistent with the fact that the pseudospin $\sigma_z$ is even
under time reversal \cite{win10}. A more quantitative discussion of
the pseudospin polarization induced by perpendicular fields
$\mathcal{E}_z$ and $B_z$ is given in Appendix~\ref{sec:polperp}.
The response of the pseudospin to the fields $\mathcal{E}_z$ and
$B_z$ is opposite to that of real spin [term (T3) in
Table~\ref{tab:invariants}], where $B_z s_z \tau_0$ induces a
real-spin polarization $\braket{s_z}$ (when averaging over the
occupied states in both valleys) while the term $\Ee_z s_z \tau_z$
does not, consistent with time reversal symmetry.  Nonetheless, both
the magnetic and the electric version of the term (T3) can be probed
experimentally via, e.g., electron spin resonance, which does not
distinguish between parallel and antiparallel spin polarizations in
the valleys $\vek{K}$ and $\vek{K}'$.

Similar to term (T3), according to term (T7), an in-plane magnetic field
$\vek{B}_\|$ gives rise to a macroscopic in-plane polarization of
real spins whereas the real-spin polarization induced by an electric
field $\vekc{E}_\|$ is anti-parallel in the two valleys. Fields
$\vek{B}_\|$ and $\vekc{E}_\|$ also couple to the in-plane
pseudospin $\sigma_\|$.  Term (T2) of Table~\ref{tab:invariants}
induces an out-of-plane tilt of the spin orientation of individual
states which, in the $\vek{K}'$ valley, has opposite signs for
$\vek{B}_\|$ and $\vekc{E}_\|$. Remarkably, on average this yields
an in-plane polarization $\braket{\vek{\sigma}_\|}$ which is
nonetheless the same in each valley for fields $\vek{B}_\|$ and
$\vekc{E}_\|$ (Fig.~\ref{fig:visual}), see Appendix~\ref{sec:polpar}
for a more quantitative discussion.  This result reflects the fact
that the macroscopic pseudospin polarization
$\braket{\vek{\sigma}_\|}$ is neither even nor odd under time
reversal \cite{win10}. More precisely, in each valley the direction
of $\braket{\vek{\sigma}_\|}$ is well-defined only up to a
gauge-dependent angular offset. Yet the \emph{change} of
$\braket{\vek{\sigma}_\|}$ induced by a change in the in-plane
orientation of the applied field \emph{is} well-defined and it
points clockwise in one valley and counterclockwise in the other
valley (for both $\vek{B}_\|$ and $\vekc{E}_\|$ and all terms in
Tab.~\ref{tab:invariants} giving rise to an in-plane pseudospin
orientation of individual states). Specifically for the term (T2),
if we change the in-plane orientation of the fields $\vek{B}_\|$ or
$\vekc{E}_\|$ by an angle $\varphi$, this changes the resulting
average polarization $\braket{\vek{\sigma}_\|}$ by $\pm 2\varphi$.
This implies, in particular, that reverting the direction of the
external field yields the same orientation of the induced pseudospin
polarization (see Appendix~\ref{sec:polpar}).  We see here that the
pseudospin polarization induced by external fields $\vekc{E}$ and
$\vek{B}$ behaves qualitatively different from the polarization of
real spins.  We emphasize that the measurable thermodynamic response
due to the term (T2) (see previous subsection) is not affected by
these ambiguities concerning the field-induced pseudospin textures.

The real spins also respond to electric and magnetic fields in
unusual ways.  Normally, a Zeeman term orients the spins parallel
(or antiparallel) to the applied magnetic field. According to terms
(T4), (T5), and (T6) a perpendicular magnetic field $B_z$ orients
the spins in-plane and according to term (T8), an in-plane magnetic
field can orient the spins out-of-plane.  The Zeeman terms (T4),
(T5), (T6), (T8) and (T9) orient the individual spins, but on
average these terms do not give rise to a spin polarization (not
even in individual valleys).  Thus we have here no simple relation
between the spin polarization and the thermodynamic magnetization.
This is due to the fact that ultimately these terms are caused by
spin-orbit coupling so that spin and orbital contributions to the
magnetization (and dielectric polarization) cannot be discussed
separately.

\section{Invariant Expansion for the SWM model of BLG}
\label{sec:invar-swm}

A unified picture of the magnitude of the various effects discussed
above can be derived from the multiband SWM model of BLG.  Indeed,
the SWM model takes a similar role for BLG as the well-known
multiband Kane model \cite{kan66, win03} for many zinc blende
semiconductors, where L\"owdin partitioning \cite{loe51} can be used
to predict the magnitude of effects like Zeeman splitting and Rashba
spin-orbit coupling.

The theory of invariants (Sec.~\ref{sec:invExp}) readily reproduces
the $\kk$-dependent terms in the SWM Hamiltonian (\ref{eq:tb-hams}).
An external field $\Ee_z$ results in the spin-independent and
$\kk$-independent terms
\begin{equation}
  \label{eq:swmc-ez}
  \mathcal{H}^\vek{K}_\Ee = \tvek[cccc]{
  0 & \varepsilon_{12} \Ee_z & 0 & 0 \\
  \varepsilon_{12} \Ee_z & 0 & 0 & 0 \\
  0 & 0 & \varepsilon_{33} \Ee_z & 0 \\
  0 & 0 & 0 & - \varepsilon_{33} \Ee_z } .
\end{equation}
Recent experiments \cite{zha09} have demonstrated that a
displacement field of $\sim 1$~V/nm generates a band gap of
$\sim 0.1$~eV in BLG, corresponding to
$\varepsilon_{33} \sim 0.05~e\,\mathrm{nm}$ or [see Eq.\
(\ref{eq:fieldHam})]
\begin{equation}
  g_\mathrm{e} = \frac{2 \, c \, \varepsilon_{33}}{\mu_B} \simeq 500 \quad.
\end{equation}
This value is reasonably close to the expected potential difference
induced by a field $\Ee_z$ between the $B$ and $B'$ layer of BLG
with an inter-layer separation $0.34$~nm that gets effectively
reduced by screening.  We may expect a similar potential difference
induced between the layers $A$ and $A'$, implying
$\varepsilon_{12} \simeq \varepsilon_{33}$.  This is also consistent
with the ab-initio calculations in Ref.~\onlinecite{kon12} that
found $\varepsilon_{12} = \varepsilon_{33} = 0.048~e\,\mathrm{nm}$.

It follows from Tables~\ref{tab:basemat} and \ref{tab:tensorop} and
Eq.\ (\ref{eq:timea2}) that intrinsic (Pauli) spin-orbit coupling
results in the terms
\begin{equation}
  \label{eq:swmc-pauli}
  \mathcal{H}^\vek{K}_p = \tvek[cccc]{
  0 & p_{12} s_z & 2i p_{13} s_- &  2i p_{13} s_+ \\
  p_{12} s_z & 0 & 2i p_{23} s_- & -2i p_{23} s_+ \\
  -2i p_{13} s_+ & -2i p_{23} s_+ & p_{33} s_z & 0 \\
  -2i p_{13} s_- &  2i p_{23} s_- & 0 & - p_{33} s_z }.
\end{equation}
Rashba spin-orbit coupling results in the terms
\begin{equation}
  \label{eq:swmc-rashba}
  \mathcal{H}^\vek{K}_r = \Ee_z \tvek[cccc]{
  r_{11} s_z & 0 & 2i r_{13} s_- & -2i r_{13} s_+ \\
  0 & r_{22} s_z & 2i r_{23} s_- & 2i r_{23} s_+ \\
  -2i r_{13} s_+ & -2i r_{23} s_+ & r_{33}' s_z & i r_{33} s_- \\
  2i r_{13} s_- & -2i r_{23} s_- & -i r_{33} s_+ &  r_{33}' s_z }.
\end{equation}
Here we have used the phase convention that all prefactors are real.

The magnitude of spin-orbit coupling in BLG was recently studied by
Konschuh \etal\ (Ref.~\onlinecite{kon12}), who developed a
generalized SWM-like tight-binding model that they fitted to the
results of numerical ab-initio calculations.  To compare their work
with ours, we need to apply the unitary transformation $U^{-1} H U$
to Eqs.\ (2) and (8) of Ref.~\onlinecite{kon12} with
\begin{equation}
  \label{eq:fabian-u}
  U = \tvek[cccc]{\frac{1}{\sqrt{2}} & \frac{1}{\sqrt{2}} & 0 & 0\\[1ex]
      0 & 0 & -1 & 0 \\[1ex] 0 & 0 & 0 & -1 \\[1ex]
      \frac{1}{\sqrt{2}} & -\frac{1}{\sqrt{2}} & 0 & 0 }.
\end{equation}
The orbital TB Hamiltonian (2) in Ref.~\onlinecite{kon12} is then
identical in form to our Eq.\ (\ref{eq:tb-hamp}), except for the
terms proportional to $\gamma_4$ which have the opposite sign.  The
unitary transformation of the SO Hamiltonian (8) in
Ref.~\onlinecite{kon12} yields
\begin{widetext}
\begin{equation}
  \label{eq:fabian-so}
  \begin{array}[b]{rl}
  U^{-1} H_\mathrm{SO} U = \quad & \tvek[*{4}{>{\Ds}c}]{
    \frac{\lambda_{\mathrm{I}2} - \lambda_{\mathrm{I}2}'}{2} s_z
  & \frac{\lambda_{\mathrm{I}2} + \lambda_{\mathrm{I}2}'}{2} s_z
  & - i \frac{\bar{\lambda}_0 + \bar{\lambda}_4}{\sqrt{2}} s_-
  & - i \frac{\bar{\lambda}_0 + \bar{\lambda}_4}{\sqrt{2}} s_+ \\
    \frac{\lambda_{\mathrm{I}2} + \lambda_{\mathrm{I}2}'}{2} s_z
  & \frac{\lambda_{\mathrm{I}2} - \lambda_{\mathrm{I}2}'}{2} s_z
  & - i \frac{\bar{\lambda}_0 - \bar{\lambda}_4}{\sqrt{2}} s_-
  &   i \frac{\bar{\lambda}_0 - \bar{\lambda}_4}{\sqrt{2}} s_+ \\
      i \frac{\bar{\lambda}_0 + \bar{\lambda}_4}{\sqrt{2}} s_+
  &   i \frac{\bar{\lambda}_0 - \bar{\lambda}_4}{\sqrt{2}} s_+
  & - \lambda_{\mathrm{I}1} s_z & 0 \\
      i \frac{\bar{\lambda}_0 + \bar{\lambda}_4}{\sqrt{2}} s_- 
  & - i \frac{\bar{\lambda}_0 - \bar{\lambda}_4}{\sqrt{2}} s_- 
  & 0 & \lambda_{\mathrm{I}1}' s_z } \\ \\ +
  & \tvek[*{4}{>{\Ds}c}]{
  \lambda_1 s_z & 0 
  & - i \frac{2 \lambda_\mathrm{BR} - \delta\lambda_4}{\sqrt{2}} s_-
  &   i \frac{2 \lambda_\mathrm{BR} - \delta\lambda_4}{\sqrt{2}} s_+ \\
  0 & - \lambda_1 s_z 
  & - i \frac{2 \lambda_\mathrm{BR} + \delta\lambda_4}{\sqrt{2}} s_-
  & - i \frac{2 \lambda_\mathrm{BR} + \delta\lambda_4}{\sqrt{2}} s_+ \\
      i \frac{2 \lambda_\mathrm{BR} - \delta\lambda_4}{\sqrt{2}} s_+
  &   i \frac{2 \lambda_\mathrm{BR} + \delta\lambda_4}{\sqrt{2}} s_+
  & 0 & i \lambda_3 s_- \\
    - i \frac{2 \lambda_\mathrm{BR} - \delta\lambda_4}{\sqrt{2}} s_- 
  &   i \frac{2 \lambda_\mathrm{BR} + \delta\lambda_4}{\sqrt{2}} s_-
  & - i \lambda_3 s_+ & 0 
  } ,
  \end{array}
\end{equation}
\end{widetext}
where the first matrix describes the intrinsic SO coupling and
the second matrix gives the SO coupling induced by an external
electric field $\Ee_z$.  Comparing with Eq.\ (\ref{eq:swmc-pauli}),
we thus get the correspondence
\begin{subequations}
\begin{eqnarray}
p_{12} &=& \frack{1}{2} (\lambda_{\mathrm{I}2} + \lambda_{\mathrm{I}2}'), \\
p_{13} &=& - (\bar{\lambda}_0 + \bar{\lambda}_4) / \sqrt{8}, \\
p_{23} &=& - (\bar{\lambda}_0 - \bar{\lambda}_4) / \sqrt{8}, \\
p_{33} &=& - (\lambda_{\mathrm{I}1} + \lambda_{\mathrm{I}1}') / 2.
\end{eqnarray}
\end{subequations}
By symmetry, we must have
$\lambda_{\mathrm{I}1} = \lambda_{\mathrm{I}1}'$ and
$\lambda_{\mathrm{I}2} = \lambda_{\mathrm{I}2}'$.  For the SO terms
induced by an electric field $\Ee_z$ [Eq.\ (\ref{eq:swmc-rashba})]
we get the correspondence
\begin{subequations}
\begin{eqnarray}
r_{11} \,\Ee_z & = & - r_{22} \,\Ee_z = \lambda_1, \\
r_{13} \,\Ee_z & = & - (\lambda_\mathrm{BR} - \delta\lambda_4/2) / \sqrt{2}, \\
r_{23} \,\Ee_z & = & - (\lambda_\mathrm{BR} + \delta\lambda_4/2) / \sqrt{2}, \\
r_{33} \,\Ee_z & = & \lambda_3.
\end{eqnarray}
\end{subequations}
The term proportional to $r_{33}'$ is absent in
Ref.~\onlinecite{kon12}.

A least-square fit of the numerical ab-initio results in
Ref.~\onlinecite{kon12} to the intrinsic SO Hamiltonian
(\ref{eq:swmc-pauli}) indicates (consistent with the findings in
Ref.~\onlinecite{kon12}) that all four coefficients $p_{ij}$
contribute significantly to the intrinsic SO coupling
\begin{subequations}
  \begin{align}
    p_{12} & = 10.8~\mathrm{\mu eV}, &
    p_{33} & = - 12.1~\mathrm{\mu eV}, \\
    p_{13} & = 4.3~\mathrm{\mu eV} , &
    p_{23} & = - 4.8~\mathrm{\mu eV}.
  \end{align}
\end{subequations}
On the other hand, we find that Rashba SO coupling in the
$4\times 4$ SWM model [Eq.\ (\ref{eq:swmc-rashba})] is dominated by
the coefficients
\begin{equation}
  r_{13} = - 5.0~e\,\mathrm{fm} \;, \hspace{3em}
  r_{23} = - 1.6~e\,\mathrm{fm} \quad .
\end{equation}
The coefficients $r_{11}$, $r_{22}$, $r_{33}$, and $r_{33}'$ in the
diagonal blocks are found to contribute only marginally to Rashba SO
coupling according to a comparison with the ab-initio results in
Ref.~\onlinecite{kon12}.

\section{L\"owdin partitioning for the SWM model of BLG}
\label{sec:loewdin}

L\"owdin partitioning \cite{bir74, win03} can provide explicit expressions
for the prefactors of many terms discussed in this work.  Here the
starting point is the $4\times 4$ SWM Hamiltonian (\ref{eq:tb-hams})
complemented by the field term (\ref{eq:swmc-ez}), the Pauli
Hamiltonian (\ref{eq:swmc-pauli}), the Rashba SO Hamiltonian
(\ref{eq:swmc-rashba}) as well as the potential due to an external
electric field $\vekc{E}$;
\begin{equation}
  \label{eq:SWM-full}
  \mathcal{H}^\vek{K} = \mathcal{H}^\vek{K}_\kk
  + \mathcal{H}^\vek{K}_\Ee
  + \mathcal{H}^\vek{K}_p + \mathcal{H}^\vek{K}_r
  + \vekc{E} \cdot \vek{r}\; .
\end{equation}
The $4\times 4$ model $\mathcal{H}^\vek{K}$ provides a comprehensive
description of the electron dynamics in BLG in the presence of
electric and magnetic fields as well as intrinsic and extrinsic SO
coupling.  Zeeman-like terms proportional to $B_z$ arise in L\"owdin
partitioning due to the noncommutativity of the components of
kinetic crystal momentum $\hbar\kk$, where
$[k_x, k_y] = (e/i\hbar)B_z$.  Similarly, Rashba-like terms
proportional to an in-plane electric field
$\vekc{E}_\| = (\Ee_x, \Ee_y, 0)$ arise from the commutator between
the last term in Eq.\ (\ref{eq:SWM-full}) and the operator $\kk$
\cite{win03, abinitio}.

As noted in the previous section, a least-square fit to the
ab-initio calculations in Ref.~\onlinecite{kon12} cannot provide
reliable values for the prefactors $r_{11}$, $r_{22}$, $r_{33}$, and
$r_{33}'$ in $\mathcal{H}^\vek{K}_r$.  However, if we project the
$4\times 4$ Hamiltonian (\ref{eq:SWM-full}) on the subspaces
$(\mathcal{H}^\vek{K})_{11}$, $(\mathcal{H}^\vek{K})_{22}$, and
$(\mathcal{H}^\vek{K})_{33}$, we obtain in lowest order
\begin{subequations}
  \begin{eqnarray}
    \tilde{r}_{11} &=&   \frac{\varepsilon_{12} \, p_{12}}{\gamma_1}
                       - \frac{8 p_{13} \, r_{13}}{\gamma_1},\\
    \tilde{r}_{22} &=& - \frac{\varepsilon_{12} \, p_{12}}{\gamma_1}
                       + \frac{8 p_{23} \, r_{23}}{\gamma_1},\\
    \tilde{r}_{33} &=& 0, \hspace{2em}
    \tilde{r}_{33}' = \frac{4 (p_{23} \, r_{23} - p_{13} \, r_{13})}{\gamma_1}
  \end{eqnarray}
\end{subequations}
where the tilde indicates that these expressions are valid only
within the respective projected subspaces.  These formulas indicate
that the $k$-independent Rashba spin splitting in the subspaces
$(\mathcal{H}^\vek{K})_{11}$ and $(\mathcal{H}^\vek{K})_{22}$ is of
first order in SO coupling whereas it is of second order in the
$(\mathcal{H}^\vek{K})_{33}$ subspace, consistent with the numerical
ab-initio calculations in Ref.~\onlinecite{kon12}.

Most often we are interested only in the subspace
$(\mathcal{H}^\vek{K})_{33}$.  In Table~\ref{tab:prefact} we
summarize the most important terms appearing in the L\"owdin
projection of $\mathcal{H}^\vek{K}$ onto the subspace
$(\mathcal{H}^\vek{K})_{33}$.

\begin{table*}
  \caption{\label{tab:prefact} Prefactors of selected terms in
  $\mathcal{H}_{33}$, obtained via L\"owdin partitioning of the
  SWM model. Both the expressions in terms of SWM parameters as well
  as the numerical values are given. The place in this article where
  a term has been introduced is also referenced.  In the first
  column, the upper sign holds for valley $\vek{K}$, the lower sign
  for $\vek{K}'$.}

\vspace{1ex}
\renewcommand{\arraystretch}{1.85}
\begin{tabular}{Ls{2em}>{$\Ds}l<{$}s{2em}Ls{2em}r@{}} \hline\hline
  \mbox{term} & \mbox{prefactor (parametric)} & \mbox{prefactor (numeric)}
  & introduced in \\ \hline
 - (k_- \, \sigma_\pm + k_+ \, \sigma_\mp) \, \tau_z &
  \tgamma_{31} & \num{0.053}{eV^2 \, nm} & (\ref{eq:eff_BLG}) \\
  k^2 \, \tau_0  &
  \tgamma_0' + \frac{2\tgamma_0\tgamma_4}{\gamma_1} &
  \num{0.13}{eV \, nm^2} & (\ref{eq:eff_BLG}) \\[2ex]
  - (k_+^2 \, \sigma_\pm + k_-^2 \, \sigma_\mp) \, \tau_0 &
  - \tgamma_{32} + \frac{\tgamma_0^2 + \tgamma_4^2}{\gamma_1}
  & \num{1.3}{eV \, nm^2} & (\ref{eq:eff_BLG}) \\[2ex] \hline
  s_z \sigma_z \tau_z & p_{33} & \num{-12}{\mu eV}
  & (\ref{eq:bia-zero}) \\[1ex]
  i(s_+ k_- - s_- k_+) \sigma_z \, \tau_z &
  \frac{\sqrt{2} \, [\tgamma_0 (p_{13} - p_{23}) 
                - \tgamma_4 (p_{13} + p_{23})]}{\gamma_1}
  & \num{18}{\mu eV \, nm} & (\ref{eq:bia-lin}) \\[2ex] \hline \rule{0pt}{4.0ex}
  B_z \sigma_z \, \tau_z &
  - \frac{2\tgamma_0\tgamma_4}{\gamma_1} \frac{e}{\hbar} &
  \num{-180}{\mu eV / T} & (T1) \\
  \Ee_z \sigma_z \, \tau_0 & \varepsilon_{33}
  & 0.048~e \, \mathrm{nm} & (T1) \\
  \Ee_z B_z \, \tau_z  &
  \frac{\tgamma_0^2 (\varepsilon_{33} + \varepsilon_{12})
      + \tgamma_4^2 (\varepsilon_{33} - \varepsilon_{12})}{\gamma_1^2} \,
  \frac{e}{\hbar} & \nume[e\,]{5.8}{-4}{nm/T} & (\ref{eq:fieldHam}) \\[2ex]
  \hline \rule{0pt}{4.5ex}
  (i/2)(k_+ \Ee_- - k_- \Ee_+) \sigma_z \, \tau_z &
  \frac{(\tgamma_0^2 + \tgamma_4^2) \, e}{\gamma_1^2} &
  \num[e\,]{4.0}{nm^2} & (T2) \\[2ex] \hline \rule{0pt}{4.5ex}
  B_z s_z \, \tau_0  &
  \frac{\tgamma_0^2 (p_{33} + p_{12}) + \tgamma_4^2 (p_{33} - p_{12})}
       {\gamma_1^2} \frac{e}{\hbar} & \nume{-9.3}{-9}{eV/T} & (T3) \\
  \Ee_z s_z \, \tau_z  &
  r_{33}' + \frac{4 (p_{23} \, r_{23} - p_{13} \, r_{13})}{\gamma_1}
  & \nume[e\,]{3.6}{-10}{nm} & (T3) \\
  \Ee_z B_z s_z \sigma_z \, \tau_0 &
  \frac{r_{33}' (\tgamma_0^2 + \tgamma_4^2)}{\gamma_1^2} \frac{e}{\hbar}
  & \approx 0 & \\[1ex] \hline \rule{0pt}{4.5ex}
  2i B_z (s_- \sigma_\pm - s_+ \sigma_\mp) \, \tau_0 &
  - \frac{\tgamma_{31} [\tgamma_0 (p_{23} + p_{13})
                      + \tgamma_4 (p_{23} - p_{13})]}{\sqrt{2} \, \gamma_1^2}
  \frac{e}{\hbar} & \nume{3.3}{-10}{eV/T} & (T4) \\[1ex]
  2i \Ee_z (s_- \sigma_\pm - s_+ \sigma_\mp) \, \tau_z & r_{33} &
  \approx 0 & (T4) \\
  \hline \rule{0pt}{4.5ex}
  i B_z (s_+ k_- - s_- k_+) \, \tau_z  &
  \frac{\sqrt{2} \, \tgamma_0'[\tgamma_0 (p_{23} + p_{13})
                            + \tgamma_4 (p_{23} - p_{13})]}{\gamma_1^2}
  \frac{e}{\hbar} & \nume{1.6}{-7}{eV\, nm/T}& (T5) \\
  i \Ee_z (s_+ k_- - s_- k_+) \, \tau_0  &
  \frac{\sqrt{2} \, [\tgamma_0 (r_{13} - r_{23})
                - \tgamma_4 (r_{13} + r_{23})]}{\gamma_1} &
  \nume[e\,]{8.7}{-6}{nm^2}& (T5) \\
  2i B_z (s_- k_- \sigma_\mp - s_+ k_+ \sigma_\pm) \, \tau_z &
  \frac{\sqrt{2}}{\gamma_1^2}
  \begin{array}[t]{@{}r@{}l@{}}
    \{ & \tgamma_0' [\tgamma_0 (p_{13} - p_{23})
                     - \tgamma_4 (p_{13} + p_{23})] \\
     \makebox[0pt][r]{${}+{}$} & 
         \tgamma_{32} [\tgamma_4 (p_{13} - p_{23})
                       - \tgamma_0 (p_{13} + p_{23})] \} \Ds\frac{e}{\hbar}
  \end{array} & \nume{-2.9}{-8}{eV\, nm/T} & (T6) \\
  2i \Ee_z (s_- k_- \sigma_\mp - s_+ k_+ \sigma_\pm) \, \tau_0 &
  \frac{\sqrt{2} \, [\tgamma_4 (r_{13} - r_{23})
                - \tgamma_0 (r_{13} + r_{23})]}{\gamma_1}
  & \nume[e\,]{2.2}{-5}{nm^2} & (T6) \\[2ex] \hline \rule{0pt}{4.5ex}
  (\Ee_+ s_- + \Ee_- s_+) \, \tau_z &
  \frac{[\tgamma_0 (p_{23} + p_{13}) + \tgamma_4 (p_{23} - p_{13})] \, e}
  {\sqrt{2}\,\gamma_1^2} & \nume[e\,]{-4.1}{-6}{nm} & (T7) \\[1ex]
  i (\Ee_- \sigma_\pm - \Ee_+ \sigma_\mp) s_z \, \tau_z & 0 & 0 & (T8) \\[1ex]
  -2 (\Ee_+ s_+ \sigma_\pm + \Ee_- s_- \sigma_\mp) \, \tau_z &
  \frac{[\tgamma_0 (p_{23} - p_{13}) + \tgamma_4 (p_{23} + p_{13})] \, e}
  {\sqrt{2}\,\gamma_1^2} & \nume[e\,]{-4.0}{-6}{nm} & (T9) \\[2ex]\hline\hline
\end{tabular}
\end{table*}

Intrinsic spin splitting in BLG is dominated by two terms, the
$k$-independent term
\begin{subequations}
  \begin{equation}\label{eq:bia-zero}
    s_z \sigma_z \,\tau_z
  \end{equation}
  orienting the spins out-of-plane as well as the $k$-linear term
  \begin{equation}\label{eq:bia-lin}
    i \left( s_+ k_- - s_- k_+ \right) \sigma_z \,\tau_0
  \end{equation}
\end{subequations}
orienting the spins in-plane.  For typical Fermi wave vectors these
terms are similar in magnitude.  The combined effect of both terms
is a vortex-like spin texture in the $(k_x,k_y)$ plane around the
$\vek{K}$ point.  Interestingly, we get rather similar spin textures
from a magnetic field $B_z$ or an electric field $\Ee_z$ via the
terms (T3), (T5) and (T6) though in each case the ratio between the
respective pre\-factors is different.

It is helpful to express the prefactors of the magnetic terms (T1),
(T3), and (T4) in terms of effective $g$ factors, giving the numeric
values $6.2$, $3.2 \times 10^{-4}$, and $1.1 \times 10^{-5}$,
respectively.  The latter two values represent actually the
correction to the $g$ factor $g=2$ of free electrons, similar to the
well-known Roth formula \cite{rot59}. Also, we may compare the
magnitude of the couplings for the pairs of electric and magnetic
terms in Table \ref{tab:invariants}, noting that for an electric
field $\Ee$ the term $\Ee/c$ has the same dimension tesla as a
magnetic field $B$.  For the term (T1), this yields
$g_\mathrm{e} / g_\mathrm{m} \approx 80$, (T3) gives the ratio
$\sim 0.012$, (T5) $\sim 16$ and (T6) $\sim 230$.  The ratio of the
couplings to the electric and magnetic fields is thus not universal.

\section{Conclusions and Outlook}
\label{sec:outlook}

Our study of the electronic structure of BLG in the presence of
electric and magnetic fields has revealed that this material
exhibits an unusual \emph{magneto-electric equivalence\/}: Every
possible coupling of a spin or pseudospin component to the electric
(magnetic) field is complemented by an analogous coupling to the
magnetic (electric) field.  Such a behavior, while
counter-intuitive, is consistent with basic requirements arising
from spatial and time-reversal symmetries.  It implies that the
thermodynamic response to a magnetic field (the magnetization) takes
the same functional form as the response to an electric field (the
polarization).  Based on L\"owdin partitioning for the multiband SWM
model of the BLG band structure and using input from ab-initio and
tight-binding calculations, we have obtained numerical prefactors
for relevant coupling terms.

Our work has focused on BLG, for which the well-established SWM
model provides a convenient basis for a systematic discussion.
However, we would like to emphasize that the group-theoretical
arguments used here are valid more generally for any multi-valley
material with similar symmetries such as silicene~\cite{gei13} and
asymmetrically hydrogenated single-layer graphene~\cite{gmi13}.
The recent interest in topological insulators has also stimulated a
quest for novel layered materials with strong spin-orbit coupling such
as Bi$_2$Se$_3$ and Sb$_2$Te$_3$, where individual layers form
a hexagonal structure similar to BLG \cite{yan12a} and therefore can
be expected to have similar electrodynamic properties.

Besides shedding new light on emergent electromagnetism in
unconventional materials, our discussion of unusual couplings
between electric and magnetic fields to spins and pseudospins due to
magneto-electric equivalence also opens up new possibilities for
creating spintronic and valleytronic devices.


\acknowledgments The authors thank M.~Meyer and H.~Schulthei{\ss}
for help with generating Fig.~\ref{fig:visual}. Useful discussions
with J.~Fabian, J.~J.~Heremans, A.~H.~MacDonald, J.~L.\ Ma\~nes, and
M.~Mor\-gen\-stern are also gratefully acknowledged. This work was
supported by the NSF under grant no.\ DMR-1310199, by Marsden Fund
contract no.\ VUW0719, administered by the Royal Society of New
Zealand, and at Argonne National Laboratory by the DOE BES via
contract no.\ DE-AC02-06-CH11357.

\appendix*
\section{Pseudospin Polarization}

In this appendix, we study more quantitatively the pseudospin
polarization induced by external fields $\vekc{E}$ and $\vek{B}$.
In the following we denote these fields by the generic
placeholder~$\vekc{F}$.

\subsection{BLG in perpendicular fields $\vekc{F}_z$}
\label{sec:polperp}

We consider the BLG Hamiltonian with term (T1) from
Table~\ref{tab:invariants}, using the simplified notation
\begin{equation}
  \mathcal{H} = a \left(k_+^2 \, \sigma_+ + k_-^2 \, \sigma_- \right)
  + c_{f,z} \mathcal{F}_z \sigma_z \;.
\end{equation}
Writing the wave vector in polar coordinates $\kk = (k,\phi)$, the
Hamiltonian becomes
\begin{equation}
  \mathcal{H} = ak^2 \tvek[cc]{ b_z & - e^{2i\phi} \\[1ex]
  - e^{-2i\phi} & - b_z} \;,
\end{equation}
with $b_z \equiv c_{f,z}\mathcal{F}_z / (ak^2)$. The pseudospin
orientation of the eigenstates $\ket{\kk}_\pm$ is
$\braket{\vek{\sigma}} (\kk,\pm) \equiv \braket{\kk | \vek{\sigma} |
\kk}_\pm$, giving
\begin{subequations}
  \label{eq:spinpol-comp}
  \begin{eqnarray}
    \braket{\sigma_x} (\kk,\pm) 
    & = & \mp \frac{\cos(2\phi)}{\sqrt{1+b_z^2}} \;, \\
    \braket{\sigma_y} (\kk,\pm) 
    & = & \pm \frac{\sin(2\phi)}{\sqrt{1+b_z^2}} \;, \\
    \braket{\sigma_z} (\kk,\pm)
    & = & \pm \frac{b}{\sqrt{1+b_z^2}} \;.
  \end{eqnarray}
\end{subequations}
The average pseudospin polarization at $T=0$ becomes
\begin{equation}
  \label{eq:spinpol-gen}
  \braket{\vek{\sigma}}_\pm 
  = \frac{2}{k_\mathrm{F}^2} \int_0^{k_\mathrm{F}} dk\, k
  \int_0^{2\pi} \frac{d\phi}{2\pi} \;
  \braket{\kk | \vek{\sigma} | \kk}_\pm  \;,
\end{equation}
giving
\begin{subequations}
  \begin{eqnarray}
    \braket{\sigma_x}_\pm & = & \braket{\sigma_y}_\pm = 0 \;, \\
    \braket{\sigma_z}_\pm & = & \pm b_{\mathrm{F},z} \,
     \ln \frac{1 + \sqrt{1 + b_{\mathrm{F},z}^2}}{b_{\mathrm{F},z}} \;,
  \end{eqnarray}
\end{subequations}
with $b_{\mathrm{F},z} \equiv c_{f,z}\mathcal{F}_z / (a
k_\mathrm{F}^2)$. We can express the average polarization
$\braket{\sigma_z}_\pm$ in terms of the Fermi energy $E_\mathrm{F}
= a k_\mathrm{F}^2 \, \sqrt{1 + b_{\mathrm{F},z}^2}$ as
\begin{equation}
  \braket{\sigma_z}_\pm  =  \pm \frac{\tilde{b}_z}{\sqrt{1-\tilde{b}_z^2}}
  \, \ln \frac{1+\sqrt{1-\tilde{b}_z^2}}{\tilde{b}_z} \quad ,
\end{equation}
with $\tilde{b}_z = c_{f,z} \mathcal{F}_z / E_\mathrm{F}$.  As to be
expected, we only have a nonzero $z$ component of the pseudospin
polarization, which changes sign when the sign of $b$ is changed.
We emphasize that the above formulas are valid for both magnetic and
electric fields, which illustrates nicely the concept of
magneto-electric equivalence.

We have $a = \hbar^2 u/(2m_0) \approx 1.27$~eV nm$^2$ and a typical
density is $n \simeq 2.5 \times 10^{12}$~cm$^{-2}$
(Ref.~\onlinecite{nov06}) giving
$k_\mathrm{F} = \sqrt{\pi \, n} \simeq 0.28$~nm$^{-1}$ and
$a k_\mathrm{F}^2 \simeq 0.1$~eV. For the electric term, we have
$g_\mathrm{e} \approx 500$ or
$c_{\mathrm{e},z} = g_\mathrm{e} \, \mu_\mathrm{B} /(2c) =
\varepsilon_{33} \approx 0.05~e \, \mathrm{nm}$,
so that $\Ee_z \sim 1$~V/nm corresponds to an electric energy
$c_{\mathrm{e},z}\Ee_z \approx 50$~meV, giving
$b_{\mathrm{e,F},z} \sim 0.5$ and
$\braket{\sigma_z}_\pm = \pm 0.76$.  For the corresponding magnetic
term, we have $g_\mathrm{m} \approx 6.2$ or
$c_{\mathrm{m},z} = g_\mathrm{m} \, \mu_\mathrm{B} / 2 \approx
0.18$~meV/T,
so that a magnetic field $B_z =10$~T corresponds to a magnetic
energy $c_{\mathrm{m},z} B_z = 1.8$~meV giving
$b_{\mathrm{m,F},z} \sim 0.018$ and
$\braket{\sigma_z}_\pm = \pm 0.085$.

\subsection{BLG in parallel fields $\vekc{F}_\|$}
\label{sec:polpar}

Next we consider the BLG Hamiltonian with term (T2) from
Table~\ref{tab:invariants},
\begin{equation}
  \label{eq:ham-plane}
  \mathcal{H} = a \left(k_+^2 \, \sigma_+ + k_-^2 \, \sigma_- \right)
  + c_{f\|} \, (k_x \mathcal{F}_y - k_y \mathcal{F}_x) \sigma_z \;.
\end{equation}
We use polar coordinates $\vekc{F}_\| = (\mathcal{F}, \theta)$ giving
\begin{equation}
  \label{eq:ham-parf}
  \mathcal{H} = a k^2 \tvek[cc]{%
  - \sqrt{2}\, b_\| / k & - e^{2i\phi} \\[1ex]
    - e^{-2i\phi} & \sqrt{2}\, b_\| / k} \;,
\end{equation}
with
$b_\| = (c_{f\|} \mathcal{F} / \sqrt{2}a) \, \sin (\phi-\theta)$.
We can then immediately express the pseudospin orientation
$\braket{\vek{\sigma}} (\kk,\pm)$ of individual states similar to
Eq.\ (\ref{eq:spinpol-comp}).  To obtain the average pseudospin
polarization of all occupied states up to the Fermi energy $E_F$, we
switch from an integration over $k$ to an integration over energy
$E$ using
\begin{equation}
  k = \sqrt{\sqrt{E^2/a^2 + b_\|^4} - b_\|^2} \quad .
\end{equation}
The pseudospin orientation of individual states becomes then
\begin{subequations}
  \begin{eqnarray}
    \braket{\sigma_x} (\kk, \pm) 
    & = & \mp \frac{\sqrt{E^2 + a^2 b_\|^4} - a b_\|^2}{E}
    \, \cos (2\phi) \;, \\
    \braket{\sigma_y} (\kk, \pm) 
    & = & \pm \frac{\sqrt{E^2 + a^2 b_\|^4} - a b_\|^2}{E}
    \, \sin (2\phi) \;, \hspace{2em} \\
    \braket{\sigma_z} (\kk, \pm)
    & = & \mp \frac{\sqrt{2} b_\| \,
    \sqrt{\sqrt{E^2 + a^2 b_\|^4} - a b_\|^2}}{E} \;.
  \end{eqnarray}
\end{subequations}
Using the fact that
\begin{equation}
\int d^2 k \equiv \int d\phi \int dk \, k = \frac{1}{2 a}
\int d\phi \int dE \, \frac{E}{\sqrt{E^2 + a^2 b_\|^4}} \,\, ,
\end{equation}
and applying the proper normalization condition, we find
\begin{subequations}
  \label{eq:6-spinpol-ang}
  \begin{eqnarray}
    \braket{\sigma_x}_\pm & = &
    \mp \mathcal{P}(\tilde{b}_\|) \, \cos (2\theta) \;, \\[1ex]
    \braket{\sigma_y}_\pm & = &
    \pm  \mathcal{P}(\tilde{b}_\|) \, \sin (2\theta) \;, \\[1ex]
    \braket{\sigma_z}_\pm & = & 0 \;,
  \end{eqnarray}
\end{subequations}
with
\begin{equation}
  \label{eq:ang_int2}
   \mathcal{P}(\tilde{b}_\|) = \frac{\displaystyle
   \int_0^{2\pi} \!\!\! d\phi\, \cos(2\phi)\, \sin^2(\phi)\,
   \ln\left[\frac{\tilde{b}_\| + \sqrt{\tilde{b}_\|^2 + \sin^4(\phi)}}
                 {\sin^2(\phi)} \right]}
   {\displaystyle \pi - \int_0^{2\pi} \!\! d\phi\,
    \sqrt{\tilde{b}_\|^2 + \sin^4(\phi)}}
\end{equation}
and $\tilde{b}_\| \equiv 2 a E_\mathrm{F}/ (c_{f\|}^2 \mathcal{F}^2)$.

It follows from Eq.\ (\ref{eq:6-spinpol-ang}) that changing the
orientation of $\vekc{F}_\|$ by the angle $\theta$ \emph{clockwise}
changes $\braket{\vek{\sigma}}_\pm$ by $2\theta$
\emph{counterclockwise}. We have, therefore, no simple geometric
relation between the orientation of $\vekc{F}_\|$ and the
orientation of the pseudospin polarization $\braket{\vek{\sigma}}$.
With our phase conventions in the $\vek{K}$ valley electric and
magnetic fields couple in the same way to the pseudospin whereas
they couple oppositely in the $\vek{K}'$ valley.  However, the above
formulas imply, in particular, that the in-plane components
$\braket{\sigma_x}$ and $\braket{\sigma_y}$ of the averaged
pseudospin polarizations are independent of the sign of
$\vekc{F}_\|$ so that electric and magnetic fields result in the
same pseudospin polarization in each valley.

Finally, we note that the trivial unitary transformation
\begin{equation}
  U (\phi_0) = \tvek[cc]{e^{i\phi_0} & 0 \\ 0 & e^{-i\phi_0}}
\end{equation}
corresponding to a rotation about the pseudospin $z$ axis turns the
Hamiltonian (\ref{eq:ham-parf}) into the unitarily equivalent
Hamiltonian
\begin{subequations}
  \begin{eqnarray} 
    \mathcal{H}' (\phi_0) & = & U \mathcal{H} U^{-1} \\
    & = & ak^2 \tvek[cc]{%
    - \sqrt{2}\, b_\|' / k & - e^{2i\phi'} \\[1ex]
    - e^{-2i\phi'} & \sqrt{2}\, b_\|' / k} \;,
  \end{eqnarray}
\end{subequations}
with
$b_\|' = (c_{f\|} \mathcal{F} / \sqrt{2}a) \, \sin (\phi'-\theta')$,
$\phi' \equiv \phi - \phi_0$ and $\theta' \equiv \theta - \phi_0$.
The pseudospin polarization $\braket{\vek{\sigma}}$ averaged over
directions $\phi'$ is then likewise rotated about the $z$ axis (by
an angle $-2\phi_0$). In that sense, any in-plane pseudospin
polarization is well-defined only up to an arbitrary angular offset
$-2\phi_0$. We can merely identify the \emph{change} of the
orientation of $\braket{\vek{\sigma}}$ when we change the
orientation of $\vekc{F}_\|$.

For the electric version of Eq.\ (\ref{eq:ham-plane}) the prefactor
$c_{\mathrm{e}\|}$ is estimated in Table~\ref{tab:prefact}. However,
for metallic BLG it is difficult to apply a significant in-plane
electric field. Assuming that the ratio
$g_\mathrm{e} / g_\mathrm{m} \approx 80$ applies not only to term
(T1) involving perpendicular fields but also to (T2) which depends
on in-plane fields, we get
$c_{\mathrm{m}\|} \simeq 0.013 \,\mathrm{eV\,nm/T}$. For a Fermi
energy $E_\mathrm{F} \simeq 0.1$~eV, an in-plane magnetic field
$B_\|=10$~T yields $\tilde{b}_{\mathrm{m}\|} \simeq 15$, resulting
in a polarization magnitude
$\mathcal{P} (\tilde{b}_{\mathrm{m}\|} \simeq 15)=0.057$.


\begin{thebibliography}{41}%
\makeatletter
\providecommand \@ifxundefined [1]{%
 \@ifx{#1\undefined}
}%
\providecommand \@ifnum [1]{%
 \ifnum #1\expandafter \@firstoftwo
 \else \expandafter \@secondoftwo
 \fi
}%
\providecommand \@ifx [1]{%
 \ifx #1\expandafter \@firstoftwo
 \else \expandafter \@secondoftwo
 \fi
}%
\providecommand \natexlab [1]{#1}%
\providecommand \enquote  [1]{``#1''}%
\providecommand \bibnamefont  [1]{#1}%
\providecommand \bibfnamefont [1]{#1}%
\providecommand \citenamefont [1]{#1}%
\providecommand \href@noop [0]{\@secondoftwo}%
\providecommand \href [0]{\begingroup \@sanitize@url \@href}%
\providecommand \@href[1]{\@@startlink{#1}\@@href}%
\providecommand \@@href[1]{\endgroup#1\@@endlink}%
\providecommand \@sanitize@url [0]{\catcode `\\12\catcode `\$12\catcode
  `\&12\catcode `\#12\catcode `\^12\catcode `\_12\catcode `\%12\relax}%
\providecommand \@@startlink[1]{}%
\providecommand \@@endlink[0]{}%
\providecommand \url  [0]{\begingroup\@sanitize@url \@url }%
\providecommand \@url [1]{\endgroup\@href {#1}{\urlprefix }}%
\providecommand \urlprefix  [0]{URL }%
\providecommand \Eprint [0]{\href }%
\providecommand \doibase [0]{http://dx.doi.org/}%
\providecommand \selectlanguage [0]{\@gobble}%
\providecommand \bibinfo  [0]{\@secondoftwo}%
\providecommand \bibfield  [0]{\@secondoftwo}%
\providecommand \translation [1]{[#1]}%
\providecommand \BibitemOpen [0]{}%
\providecommand \bibitemStop [0]{}%
\providecommand \bibitemNoStop [0]{.\EOS\space}%
\providecommand \EOS [0]{\spacefactor3000\relax}%
\providecommand \BibitemShut  [1]{\csname bibitem#1\endcsname}%
\let\auto@bib@innerbib\@empty
\bibitem [{\citenamefont {Hehl}\ \emph {et~al.}(2008)\citenamefont {Hehl},
  \citenamefont {Obukhov}, \citenamefont {Rivera},\ and\ \citenamefont
  {Schmid}}]{heh08}%
  \BibitemOpen
  \bibfield  {author} {\bibinfo {author} {\bibfnamefont {F.~W.}\ \bibnamefont
  {Hehl}}, \bibinfo {author} {\bibfnamefont {Y.~N.}\ \bibnamefont {Obukhov}},
  \bibinfo {author} {\bibfnamefont {J.-P.}\ \bibnamefont {Rivera}}, \ and\
  \bibinfo {author} {\bibfnamefont {H.}~\bibnamefont {Schmid}},\ }\href@noop {}
  {\bibfield  {journal} {\bibinfo  {journal} {Phys. Lett. A}\ }\textbf
  {\bibinfo {volume} {372}},\ \bibinfo {pages} {1141} (\bibinfo {year}
  {2008})}\BibitemShut {NoStop}%
\bibitem [{\citenamefont {Essin}\ \emph {et~al.}(2009)\citenamefont {Essin},
  \citenamefont {Moore},\ and\ \citenamefont {Vanderbilt}}]{ess09}%
  \BibitemOpen
  \bibfield  {author} {\bibinfo {author} {\bibfnamefont {A.~M.}\ \bibnamefont
  {Essin}}, \bibinfo {author} {\bibfnamefont {J.~E.}\ \bibnamefont {Moore}}, \
  and\ \bibinfo {author} {\bibfnamefont {D.}~\bibnamefont {Vanderbilt}},\
  }\href@noop {} {\bibfield  {journal} {\bibinfo  {journal} {Phys. Rev. Lett.}\
  }\textbf {\bibinfo {volume} {102}},\ \bibinfo {pages} {146805} (\bibinfo
  {year} {2009})}\BibitemShut {NoStop}%
\bibitem [{\citenamefont {Essin}\ \emph {et~al.}(2010)\citenamefont {Essin},
  \citenamefont {Turner}, \citenamefont {Moore},\ and\ \citenamefont
  {Vanderbilt}}]{ess10}%
  \BibitemOpen
  \bibfield  {author} {\bibinfo {author} {\bibfnamefont {A.~M.}\ \bibnamefont
  {Essin}}, \bibinfo {author} {\bibfnamefont {A.~M.}\ \bibnamefont {Turner}},
  \bibinfo {author} {\bibfnamefont {J.~E.}\ \bibnamefont {Moore}}, \ and\
  \bibinfo {author} {\bibfnamefont {D.}~\bibnamefont {Vanderbilt}},\
  }\href@noop {} {\bibfield  {journal} {\bibinfo  {journal} {Phys. Rev. B}\
  }\textbf {\bibinfo {volume} {81}},\ \bibinfo {pages} {205104} (\bibinfo
  {year} {2010})}\BibitemShut {NoStop}%
\bibitem [{\citenamefont {Spaldin}\ and\ \citenamefont {Fiebig}(2005)}]{spa05}%
  \BibitemOpen
  \bibfield  {author} {\bibinfo {author} {\bibfnamefont {N.~A.}\ \bibnamefont
  {Spaldin}}\ and\ \bibinfo {author} {\bibfnamefont {M.}~\bibnamefont
  {Fiebig}},\ }\href@noop {} {\bibfield  {journal} {\bibinfo  {journal}
  {Science}\ }\textbf {\bibinfo {volume} {309}},\ \bibinfo {pages} {391}
  (\bibinfo {year} {2005})}\BibitemShut {NoStop}%
\bibitem [{\citenamefont {Fiebig}(2005)}]{fie05}%
  \BibitemOpen
  \bibfield  {author} {\bibinfo {author} {\bibfnamefont {M.}~\bibnamefont
  {Fiebig}},\ }\href@noop {} {\bibfield  {journal} {\bibinfo  {journal} {J.
  Phys. D}\ }\textbf {\bibinfo {volume} {38}},\ \bibinfo {pages} {R123}
  (\bibinfo {year} {2005})}\BibitemShut {NoStop}%
\bibitem [{\citenamefont {Ramesh}\ and\ \citenamefont {Spaldin}(2007)}]{ram07}%
  \BibitemOpen
  \bibfield  {author} {\bibinfo {author} {\bibfnamefont {R.}~\bibnamefont
  {Ramesh}}\ and\ \bibinfo {author} {\bibfnamefont {N.~A.}\ \bibnamefont
  {Spaldin}},\ }\href@noop {} {\bibfield  {journal} {\bibinfo  {journal} {Nat.
  Mater.}\ }\textbf {\bibinfo {volume} {6}},\ \bibinfo {pages} {21} (\bibinfo
  {year} {2007})}\BibitemShut {NoStop}%
\bibitem [{\citenamefont {Qi}\ \emph {et~al.}(2008)\citenamefont {Qi},
  \citenamefont {Hughes},\ and\ \citenamefont {Zhang}}]{qi08}%
  \BibitemOpen
  \bibfield  {author} {\bibinfo {author} {\bibfnamefont {X.-L.}\ \bibnamefont
  {Qi}}, \bibinfo {author} {\bibfnamefont {T.~L.}\ \bibnamefont {Hughes}}, \
  and\ \bibinfo {author} {\bibfnamefont {S.-C.}\ \bibnamefont {Zhang}},\
  }\href@noop {} {\bibfield  {journal} {\bibinfo  {journal} {Phys. Rev. B}\
  }\textbf {\bibinfo {volume} {78}},\ \bibinfo {pages} {195424} (\bibinfo
  {year} {2008})}\BibitemShut {NoStop}%
\bibitem [{\citenamefont {O'Dell}(1970)}]{ode70}%
  \BibitemOpen
  \bibfield  {author} {\bibinfo {author} {\bibfnamefont {T.~H.}\ \bibnamefont
  {O'Dell}},\ }\href@noop {} {\emph {\bibinfo {title} {The Electrodynamics of
  Magneto-electric Media}}}\ (\bibinfo  {publisher} {North-Holland},\ \bibinfo
  {address} {Amsterdam},\ \bibinfo {year} {1970})\BibitemShut {NoStop}%
\bibitem [{\citenamefont {Wilczek}(1987)}]{wil87}%
  \BibitemOpen
  \bibfield  {author} {\bibinfo {author} {\bibfnamefont {F.}~\bibnamefont
  {Wilczek}},\ }\href@noop {} {\bibfield  {journal} {\bibinfo  {journal} {Phys.
  Rev. Lett.}\ }\textbf {\bibinfo {volume} {58}},\ \bibinfo {pages} {1799}
  (\bibinfo {year} {1987})}\BibitemShut {NoStop}%
\bibitem [{\citenamefont {Franz}(2008)}]{fra08}%
  \BibitemOpen
  \bibfield  {author} {\bibinfo {author} {\bibfnamefont {M.}~\bibnamefont
  {Franz}},\ }\href@noop {} {\bibfield  {journal} {\bibinfo  {journal}
  {Physics}\ }\textbf {\bibinfo {volume} {1}},\ \bibinfo {pages} {36} (\bibinfo
  {year} {2008})}\BibitemShut {NoStop}%
\bibitem [{\citenamefont {Z\"ulicke}\ and\ \citenamefont
  {Winkler}(2014)}]{zue14}%
  \BibitemOpen
  \bibfield  {author} {\bibinfo {author} {\bibfnamefont {U.}~\bibnamefont
  {Z\"ulicke}}\ and\ \bibinfo {author} {\bibfnamefont {R.}~\bibnamefont
  {Winkler}},\ }\href {\doibase 10.1103/PhysRevB.90.125412} {\bibfield
  {journal} {\bibinfo  {journal} {Phys. Rev. B}\ }\textbf {\bibinfo {volume}
  {90}},\ \bibinfo {pages} {125412} (\bibinfo {year} {2014})}\BibitemShut
  {NoStop}%
\bibitem [{\citenamefont {McCann}\ and\ \citenamefont {Fal'ko}(2006)}]{mcc06}%
  \BibitemOpen
  \bibfield  {author} {\bibinfo {author} {\bibfnamefont {E.}~\bibnamefont
  {McCann}}\ and\ \bibinfo {author} {\bibfnamefont {V.~I.}\ \bibnamefont
  {Fal'ko}},\ }\href@noop {} {\bibfield  {journal} {\bibinfo  {journal} {Phys.
  Rev. Lett.}\ }\textbf {\bibinfo {volume} {96}},\ \bibinfo {eid} {086805}
  (\bibinfo {year} {2006})}\BibitemShut {NoStop}%
\bibitem [{\citenamefont {{Castro Neto}}\ \emph {et~al.}(2009)\citenamefont
  {{Castro Neto}}, \citenamefont {Guinea}, \citenamefont {Peres}, \citenamefont
  {Novoselov},\ and\ \citenamefont {Geim}}]{cas09}%
  \BibitemOpen
  \bibfield  {author} {\bibinfo {author} {\bibfnamefont {A.~H.}\ \bibnamefont
  {{Castro Neto}}}, \bibinfo {author} {\bibfnamefont {F.}~\bibnamefont
  {Guinea}}, \bibinfo {author} {\bibfnamefont {N.~M.~R.}\ \bibnamefont
  {Peres}}, \bibinfo {author} {\bibfnamefont {K.~S.}\ \bibnamefont
  {Novoselov}}, \ and\ \bibinfo {author} {\bibfnamefont {A.~K.}\ \bibnamefont
  {Geim}},\ }\href@noop {} {\bibfield  {journal} {\bibinfo  {journal} {Rev.
  Mod. Phys.}\ }\textbf {\bibinfo {volume} {81}},\ \bibinfo {pages} {109}
  (\bibinfo {year} {2009})}\BibitemShut {NoStop}%
\bibitem [{\citenamefont {McCann}\ and\ \citenamefont {Koshino}(2013)}]{mcc13}%
  \BibitemOpen
  \bibfield  {author} {\bibinfo {author} {\bibfnamefont {E.}~\bibnamefont
  {McCann}}\ and\ \bibinfo {author} {\bibfnamefont {M.}~\bibnamefont
  {Koshino}},\ }\href {http://stacks.iop.org/0034-4885/76/i=5/a=056503}
  {\bibfield  {journal} {\bibinfo  {journal} {Rep. Prog. Phys.}\ }\textbf
  {\bibinfo {volume} {76}},\ \bibinfo {pages} {056503} (\bibinfo {year}
  {2013})}\BibitemShut {NoStop}%
\bibitem [{\citenamefont {Kittel}(1963)}]{kit63}%
  \BibitemOpen
  \bibfield  {author} {\bibinfo {author} {\bibfnamefont {C.}~\bibnamefont
  {Kittel}},\ }\enquote {\bibinfo {title} {Quantum theory of solids},}\ \
  (\bibinfo  {publisher} {Wiley},\ \bibinfo {address} {New York},\ \bibinfo
  {year} {1963})\ Chap.~\bibinfo {chapter} {14}\BibitemShut {NoStop}%
\bibitem [{\citenamefont {McCann}(2006)}]{mcc06b}%
  \BibitemOpen
  \bibfield  {author} {\bibinfo {author} {\bibfnamefont {E.}~\bibnamefont
  {McCann}},\ }\href {\doibase 10.1103/PhysRevB.74.161403} {\bibfield
  {journal} {\bibinfo  {journal} {Phys. Rev. B}\ }\textbf {\bibinfo {volume}
  {74}},\ \bibinfo {pages} {161403} (\bibinfo {year} {2006})}\BibitemShut
  {NoStop}%
\bibitem [{\citenamefont {Ohta}\ \emph {et~al.}(2006)\citenamefont {Ohta},
  \citenamefont {Bostwick}, \citenamefont {Seyller}, \citenamefont {Horn},\
  and\ \citenamefont {Rotenberg}}]{oht06}%
  \BibitemOpen
  \bibfield  {author} {\bibinfo {author} {\bibfnamefont {T.}~\bibnamefont
  {Ohta}}, \bibinfo {author} {\bibfnamefont {A.}~\bibnamefont {Bostwick}},
  \bibinfo {author} {\bibfnamefont {T.}~\bibnamefont {Seyller}}, \bibinfo
  {author} {\bibfnamefont {K.}~\bibnamefont {Horn}}, \ and\ \bibinfo {author}
  {\bibfnamefont {E.}~\bibnamefont {Rotenberg}},\ }\href {\doibase
  10.1126/science.1130681} {\bibfield  {journal} {\bibinfo  {journal}
  {Science}\ }\textbf {\bibinfo {volume} {313}},\ \bibinfo {pages} {951}
  (\bibinfo {year} {2006})}\BibitemShut {NoStop}%
\bibitem [{\citenamefont {Zhang}\ \emph {et~al.}(2009)\citenamefont {Zhang},
  \citenamefont {Tang}, \citenamefont {Girit}, \citenamefont {Hao},
  \citenamefont {Martin}, \citenamefont {Zettl}, \citenamefont {Crommie},
  \citenamefont {Shen},\ and\ \citenamefont {Wang}}]{zha09}%
  \BibitemOpen
  \bibfield  {author} {\bibinfo {author} {\bibfnamefont {Y.}~\bibnamefont
  {Zhang}}, \bibinfo {author} {\bibfnamefont {T.-T.}\ \bibnamefont {Tang}},
  \bibinfo {author} {\bibfnamefont {C.}~\bibnamefont {Girit}}, \bibinfo
  {author} {\bibfnamefont {Z.}~\bibnamefont {Hao}}, \bibinfo {author}
  {\bibfnamefont {M.~C.}\ \bibnamefont {Martin}}, \bibinfo {author}
  {\bibfnamefont {A.}~\bibnamefont {Zettl}}, \bibinfo {author} {\bibfnamefont
  {M.~F.}\ \bibnamefont {Crommie}}, \bibinfo {author} {\bibfnamefont {Y.~R.}\
  \bibnamefont {Shen}}, \ and\ \bibinfo {author} {\bibfnamefont
  {F.}~\bibnamefont {Wang}},\ }\href@noop {} {\bibfield  {journal} {\bibinfo
  {journal} {Nature}\ }\textbf {\bibinfo {volume} {459}},\ \bibinfo {pages}
  {820} (\bibinfo {year} {2009})}\BibitemShut {NoStop}%
\bibitem [{\citenamefont {Zhang}\ \emph {et~al.}(2011)\citenamefont {Zhang},
  \citenamefont {Fogler},\ and\ \citenamefont {Arovas}}]{zha11}%
  \BibitemOpen
  \bibfield  {author} {\bibinfo {author} {\bibfnamefont {L.~M.}\ \bibnamefont
  {Zhang}}, \bibinfo {author} {\bibfnamefont {M.~M.}\ \bibnamefont {Fogler}}, \
  and\ \bibinfo {author} {\bibfnamefont {D.~P.}\ \bibnamefont {Arovas}},\
  }\href@noop {} {\bibfield  {journal} {\bibinfo  {journal} {Phys. Rev. B}\
  }\textbf {\bibinfo {volume} {84}},\ \bibinfo {pages} {075451} (\bibinfo
  {year} {2011})}\BibitemShut {NoStop}%
\bibitem [{\citenamefont {Xiao}\ \emph {et~al.}(2007)\citenamefont {Xiao},
  \citenamefont {Yao},\ and\ \citenamefont {Niu}}]{xia07}%
  \BibitemOpen
  \bibfield  {author} {\bibinfo {author} {\bibfnamefont {D.}~\bibnamefont
  {Xiao}}, \bibinfo {author} {\bibfnamefont {W.}~\bibnamefont {Yao}}, \ and\
  \bibinfo {author} {\bibfnamefont {Q.}~\bibnamefont {Niu}},\ }\href@noop {}
  {\bibfield  {journal} {\bibinfo  {journal} {Phys. Rev. Lett.}\ }\textbf
  {\bibinfo {volume} {99}},\ \bibinfo {pages} {236809} (\bibinfo {year}
  {2007})}\BibitemShut {NoStop}%
\bibitem [{\citenamefont {Nakamura}\ \emph {et~al.}(2009)\citenamefont
  {Nakamura}, \citenamefont {Castro},\ and\ \citenamefont {D\'ora}}]{nak09}%
  \BibitemOpen
  \bibfield  {author} {\bibinfo {author} {\bibfnamefont {M.}~\bibnamefont
  {Nakamura}}, \bibinfo {author} {\bibfnamefont {E.~V.}\ \bibnamefont
  {Castro}}, \ and\ \bibinfo {author} {\bibfnamefont {B.}~\bibnamefont
  {D\'ora}},\ }\href@noop {} {\bibfield  {journal} {\bibinfo  {journal} {Phys.
  Rev. Lett.}\ }\textbf {\bibinfo {volume} {103}},\ \bibinfo {pages} {266804}
  (\bibinfo {year} {2009})}\BibitemShut {NoStop}%
\bibitem [{\citenamefont {Koshino}\ and\ \citenamefont {McCann}(2010)}]{kos10}%
  \BibitemOpen
  \bibfield  {author} {\bibinfo {author} {\bibfnamefont {M.}~\bibnamefont
  {Koshino}}\ and\ \bibinfo {author} {\bibfnamefont {E.}~\bibnamefont
  {McCann}},\ }\href@noop {} {\bibfield  {journal} {\bibinfo  {journal} {Phys.
  Rev. B}\ }\textbf {\bibinfo {volume} {81}},\ \bibinfo {pages} {115315}
  (\bibinfo {year} {2010})}\BibitemShut {NoStop}%
\bibitem [{\citenamefont {Wallace}(1947)}]{wal47}%
  \BibitemOpen
  \bibfield  {author} {\bibinfo {author} {\bibfnamefont {P.~R.}\ \bibnamefont
  {Wallace}},\ }\href {\doibase 10.1103/PhysRev.71.622} {\bibfield  {journal}
  {\bibinfo  {journal} {Phys. Rev.}\ }\textbf {\bibinfo {volume} {71}},\
  \bibinfo {pages} {622} (\bibinfo {year} {1947})}\BibitemShut {NoStop}%
\bibitem [{\citenamefont {McClure}(1957)}]{mcc57}%
  \BibitemOpen
  \bibfield  {author} {\bibinfo {author} {\bibfnamefont {J.~W.}\ \bibnamefont
  {McClure}},\ }\href@noop {} {\bibfield  {journal} {\bibinfo  {journal} {Phys.
  Rev.}\ }\textbf {\bibinfo {volume} {108}},\ \bibinfo {pages} {612} (\bibinfo
  {year} {1957})}\BibitemShut {NoStop}%
\bibitem [{\citenamefont {Koster}\ \emph {et~al.}(1963)\citenamefont {Koster},
  \citenamefont {Dimmock}, \citenamefont {Wheeler},\ and\ \citenamefont
  {Statz}}]{kos63}%
  \BibitemOpen
  \bibfield  {author} {\bibinfo {author} {\bibfnamefont {G.~F.}\ \bibnamefont
  {Koster}}, \bibinfo {author} {\bibfnamefont {J.~O.}\ \bibnamefont {Dimmock}},
  \bibinfo {author} {\bibfnamefont {R.~G.}\ \bibnamefont {Wheeler}}, \ and\
  \bibinfo {author} {\bibfnamefont {H.}~\bibnamefont {Statz}},\ }\href@noop {}
  {\emph {\bibinfo {title} {Properties of the Thirty-Two Point Groups}}}\
  (\bibinfo  {publisher} {MIT},\ \bibinfo {address} {Cambridge, MA},\ \bibinfo
  {year} {1963})\BibitemShut {NoStop}%
\bibitem [{\citenamefont {Bir}\ and\ \citenamefont {Pikus}(1974)}]{bir74}%
  \BibitemOpen
  \bibfield  {author} {\bibinfo {author} {\bibfnamefont {G.~L.}\ \bibnamefont
  {Bir}}\ and\ \bibinfo {author} {\bibfnamefont {G.~E.}\ \bibnamefont
  {Pikus}},\ }\href@noop {} {\emph {\bibinfo {title} {Symmetry and
  Strain-Induced Effects in Semiconductors}}}\ (\bibinfo  {publisher} {Wiley},\
  \bibinfo {address} {New York},\ \bibinfo {year} {1974})\BibitemShut {NoStop}%
\bibitem [{\citenamefont {Winkler}(2003)}]{win03}%
  \BibitemOpen
  \bibfield  {author} {\bibinfo {author} {\bibfnamefont {R.}~\bibnamefont
  {Winkler}},\ }\href@noop {} {\emph {\bibinfo {title} {Spin-Orbit Coupling
  Effects in Two-Dimensional Electron and Hole Systems}}}\ (\bibinfo
  {publisher} {Springer},\ \bibinfo {address} {Berlin},\ \bibinfo {year}
  {2003})\BibitemShut {NoStop}%
\bibitem [{\citenamefont {Winkler}\ and\ \citenamefont
  {Z\"ulicke}(2010)}]{win10a}%
  \BibitemOpen
  \bibfield  {author} {\bibinfo {author} {\bibfnamefont {R.}~\bibnamefont
  {Winkler}}\ and\ \bibinfo {author} {\bibfnamefont {U.}~\bibnamefont
  {Z\"ulicke}},\ }\href {\doibase 10.1103/PhysRevB.82.245313} {\bibfield
  {journal} {\bibinfo  {journal} {Phys. Rev. B}\ }\textbf {\bibinfo {volume}
  {82}},\ \bibinfo {pages} {245313} (\bibinfo {year} {2010})}\BibitemShut
  {NoStop}%
\bibitem [{\citenamefont {Ma\~nes}\ \emph {et~al.}(2007)\citenamefont
  {Ma\~nes}, \citenamefont {Guinea},\ and\ \citenamefont {Vozmediano}}]{man07}%
  \BibitemOpen
  \bibfield  {author} {\bibinfo {author} {\bibfnamefont {J.~L.}\ \bibnamefont
  {Ma\~nes}}, \bibinfo {author} {\bibfnamefont {F.}~\bibnamefont {Guinea}}, \
  and\ \bibinfo {author} {\bibfnamefont {M.~A.~H.}\ \bibnamefont
  {Vozmediano}},\ }\href@noop {} {\bibfield  {journal} {\bibinfo  {journal}
  {Phys. Rev. B}\ }\textbf {\bibinfo {volume} {75}},\ \bibinfo {pages} {155424}
  (\bibinfo {year} {2007})}\BibitemShut {NoStop}%
\bibitem [{\citenamefont {Winkler}\ and\ \citenamefont {Z\"ulicke}()}]{win12}%
  \BibitemOpen
  \bibfield  {author} {\bibinfo {author} {\bibfnamefont {R.}~\bibnamefont
  {Winkler}}\ and\ \bibinfo {author} {\bibfnamefont {U.}~\bibnamefont
  {Z\"ulicke}},\ }\href@noop {} {}\bibinfo {note} {{p}reprint
  arXiv:1206.4761}\BibitemShut {NoStop}%
\bibitem [{\citenamefont {Geissler}\ \emph {et~al.}(2013)\citenamefont
  {Geissler}, \citenamefont {Budich},\ and\ \citenamefont
  {Trauzettel}}]{gei13}%
  \BibitemOpen
  \bibfield  {author} {\bibinfo {author} {\bibfnamefont {F.}~\bibnamefont
  {Geissler}}, \bibinfo {author} {\bibfnamefont {J.~C.}\ \bibnamefont
  {Budich}}, \ and\ \bibinfo {author} {\bibfnamefont {B.}~\bibnamefont
  {Trauzettel}},\ }\href {\doibase 10.1088/1367-2630/15/8/085030} {\bibfield
  {journal} {\bibinfo  {journal} {New J. Phys.}\ }\textbf {\bibinfo {volume}
  {15}},\ \bibinfo {pages} {085030} (\bibinfo {year} {2013})}\BibitemShut
  {NoStop}%
\bibitem [{\citenamefont {Gmitra}\ \emph {et~al.}(2013)\citenamefont {Gmitra},
  \citenamefont {Kochan},\ and\ \citenamefont {Fabian}}]{gmi13}%
  \BibitemOpen
  \bibfield  {author} {\bibinfo {author} {\bibfnamefont {M.}~\bibnamefont
  {Gmitra}}, \bibinfo {author} {\bibfnamefont {D.}~\bibnamefont {Kochan}}, \
  and\ \bibinfo {author} {\bibfnamefont {J.}~\bibnamefont {Fabian}},\ }\href
  {\doibase 10.1103/PhysRevLett.110.246602} {\bibfield  {journal} {\bibinfo
  {journal} {Phys. Rev. Lett.}\ }\textbf {\bibinfo {volume} {110}},\ \bibinfo
  {pages} {246602} (\bibinfo {year} {2013})}\BibitemShut {NoStop}%
\bibitem [{\citenamefont {Ashcroft}\ and\ \citenamefont
  {Mermin}(1976)}]{ash76}%
  \BibitemOpen
  \bibfield  {author} {\bibinfo {author} {\bibfnamefont {N.~W.}\ \bibnamefont
  {Ashcroft}}\ and\ \bibinfo {author} {\bibfnamefont {N.~D.}\ \bibnamefont
  {Mermin}},\ }\enquote {\bibinfo {title} {Solid state physics},}\ \ (\bibinfo
  {publisher} {Holt, Rinehart, Winston},\ \bibinfo {address} {Philadelphia},\
  \bibinfo {year} {1976})\ Chap.~\bibinfo {chapter} {31}\BibitemShut {NoStop}%
\bibitem [{\citenamefont {Winkler}\ and\ \citenamefont
  {Z{\"u}licke}(2010)}]{win10}%
  \BibitemOpen
  \bibfield  {author} {\bibinfo {author} {\bibfnamefont {R.}~\bibnamefont
  {Winkler}}\ and\ \bibinfo {author} {\bibfnamefont {U.}~\bibnamefont
  {Z{\"u}licke}},\ }\href@noop {} {\bibfield  {journal} {\bibinfo  {journal}
  {Phys. Lett. A}\ }\textbf {\bibinfo {volume} {374}},\ \bibinfo {pages} {4003}
  (\bibinfo {year} {2010})}\BibitemShut {NoStop}%
\bibitem [{\citenamefont {Kane}(1966)}]{kan66}%
  \BibitemOpen
  \bibfield  {author} {\bibinfo {author} {\bibfnamefont {E.~O.}\ \bibnamefont
  {Kane}},\ }in\ \href@noop {} {\emph {\bibinfo {booktitle} {Semiconductors and
  Semimetals}}},\ Vol.~\bibinfo {volume} {1},\ \bibinfo {editor} {edited by\
  \bibinfo {editor} {\bibfnamefont {R.~K.}\ \bibnamefont {Willardson}}\ and\
  \bibinfo {editor} {\bibfnamefont {A.~C.}\ \bibnamefont {Beer}}}\ (\bibinfo
  {publisher} {Academic},\ \bibinfo {address} {New York},\ \bibinfo {year}
  {1966})\ Chap.~\bibinfo {chapter} {3}, pp.\ \bibinfo {pages}
  {75--100}\BibitemShut {NoStop}%
\bibitem [{\citenamefont {L\"owdin}(1951)}]{loe51}%
  \BibitemOpen
  \bibfield  {author} {\bibinfo {author} {\bibfnamefont {P.-O.}\ \bibnamefont
  {L\"owdin}},\ }\href@noop {} {\bibfield  {journal} {\bibinfo  {journal} {J.\
  Chem.\ Phys.}\ }\textbf {\bibinfo {volume} {19}},\ \bibinfo {pages} {1396}
  (\bibinfo {year} {1951})}\BibitemShut {NoStop}%
\bibitem [{\citenamefont {Konschuh}\ \emph {et~al.}(2012)\citenamefont
  {Konschuh}, \citenamefont {Gmitra}, \citenamefont {Kochan},\ and\
  \citenamefont {Fabian}}]{kon12}%
  \BibitemOpen
  \bibfield  {author} {\bibinfo {author} {\bibfnamefont {S.}~\bibnamefont
  {Konschuh}}, \bibinfo {author} {\bibfnamefont {M.}~\bibnamefont {Gmitra}},
  \bibinfo {author} {\bibfnamefont {D.}~\bibnamefont {Kochan}}, \ and\ \bibinfo
  {author} {\bibfnamefont {J.}~\bibnamefont {Fabian}},\ }\href@noop {}
  {\bibfield  {journal} {\bibinfo  {journal} {Phys. Rev. B}\ }\textbf {\bibinfo
  {volume} {85}},\ \bibinfo {pages} {115423} (\bibinfo {year}
  {2012})}\BibitemShut {NoStop}%
\bibitem [{abi()}]{abinitio}%
  \BibitemOpen
  \href@noop {} {}\bibinfo {note} {For quasi-2D BLG the capabilities of
  ab-initio calculations and L\"owdin partitioning applied to the $4\times 4$
  SWM model are complementary: the former can incorporate an electric field
  $\Ee_z$ perpendicular to the plane of BLG and the vector potential for an
  in-plane magnetic field $\vek{B}_\parallel$ (though in
  Ref.~\onlinecite{kon12} the field $\vek{B}_\parallel$ was not considered).
  The latter can incorporate in-plane electric fields $\vekc{E}_\parallel$ and
  perpendicular magnetic fields $B_z$ (Refs.~\onlinecite{kit63,
  win03}).}\BibitemShut {Stop}%
\bibitem [{\citenamefont {Roth}\ \emph {et~al.}(1959)\citenamefont {Roth},
  \citenamefont {Lax},\ and\ \citenamefont {Zwerdling}}]{rot59}%
  \BibitemOpen
  \bibfield  {author} {\bibinfo {author} {\bibfnamefont {L.~M.}\ \bibnamefont
  {Roth}}, \bibinfo {author} {\bibfnamefont {B.}~\bibnamefont {Lax}}, \ and\
  \bibinfo {author} {\bibfnamefont {S.}~\bibnamefont {Zwerdling}},\ }\href@noop
  {} {\bibfield  {journal} {\bibinfo  {journal} {Phys. Rev.}\ }\textbf
  {\bibinfo {volume} {114}},\ \bibinfo {pages} {90} (\bibinfo {year}
  {1959})}\BibitemShut {NoStop}%
\bibitem [{\citenamefont {Yan}\ and\ \citenamefont {Zhang}(2012)}]{yan12a}%
  \BibitemOpen
  \bibfield  {author} {\bibinfo {author} {\bibfnamefont {B.}~\bibnamefont
  {Yan}}\ and\ \bibinfo {author} {\bibfnamefont {S.-C.}\ \bibnamefont
  {Zhang}},\ }\href@noop {} {\bibfield  {journal} {\bibinfo  {journal} {Rep.
  Prog. Phys.}\ }\textbf {\bibinfo {volume} {75}},\ \bibinfo {pages} {096501}
  (\bibinfo {year} {2012})}\BibitemShut {NoStop}%
\bibitem [{\citenamefont {Novoselov}\ \emph {et~al.}(2006)\citenamefont
  {Novoselov}, \citenamefont {McCann}, \citenamefont {Morozov}, \citenamefont
  {Fal'ko}, \citenamefont {Katsnelson}, \citenamefont {Zeitler}, \citenamefont
  {Jiang}, \citenamefont {Schedin},\ and\ \citenamefont {Geim}}]{nov06}%
  \BibitemOpen
  \bibfield  {author} {\bibinfo {author} {\bibfnamefont {K.~S.}\ \bibnamefont
  {Novoselov}}, \bibinfo {author} {\bibfnamefont {E.}~\bibnamefont {McCann}},
  \bibinfo {author} {\bibfnamefont {S.~V.}\ \bibnamefont {Morozov}}, \bibinfo
  {author} {\bibfnamefont {V.~I.}\ \bibnamefont {Fal'ko}}, \bibinfo {author}
  {\bibfnamefont {M.~I.}\ \bibnamefont {Katsnelson}}, \bibinfo {author}
  {\bibfnamefont {U.}~\bibnamefont {Zeitler}}, \bibinfo {author} {\bibfnamefont
  {D.}~\bibnamefont {Jiang}}, \bibinfo {author} {\bibfnamefont
  {F.}~\bibnamefont {Schedin}}, \ and\ \bibinfo {author} {\bibfnamefont
  {A.~K.}\ \bibnamefont {Geim}},\ }\href@noop {} {\bibfield  {journal}
  {\bibinfo  {journal} {Nat. Phys.}\ }\textbf {\bibinfo {volume} {2}},\
  \bibinfo {pages} {177} (\bibinfo {year} {2006})}\BibitemShut {NoStop}%
\end{thebibliography}
\end{document}